\documentclass[useAMS,usenatbib]{mn2e}
\usepackage{graphicx}
\usepackage{natbib}
\usepackage{amssymb}
\usepackage{grffile}
\usepackage{longtable,lscape}
\usepackage{color}
\usepackage{hyperref}
\usepackage{pdfpages}
\usepackage{cleveref}[2012/02/15]
\newcommand{\changes}[1] {{\color{black} #1}}
\newcommand{\changestwo}[1] {{\color{black} #1}}
\newcommand{\changesthree}[1] {{\color{black} #1}}

\title[Optical \& SZ Observations of Galaxy Clusters]
  {Optical and Sunyaev-Zel'dovich Observations of a New Sample of Distant Rich Galaxy Clusters in the 
        \textit{ROSAT} All Sky Survey}
\author[A. Buddendiek et al.]
  {A.~Buddendiek,$^{1}$\thanks{abuddend@astro.uni-bonn.de} 
  T.~Schrabback,$^{1,2,3}$
  C.~H.~Greer,$^4$
  H.~Hoekstra,$^2$
  M.~Sommer,$^1$
  \newauthor
  T.~Eifler,$^{5,6}$
  T.~Erben,$^{1}$
  J.~Erler,$^1$
  A.~K.~Hicks,$^{7}$ 
  F.~W.~High,$^{8}$ 
  H.~Hildebrandt,$^{1}$
  \newauthor
  D.~P.~Marrone,$^{4}$
  R.~G.~Morris,$^{3,9}$ 
  A.~Muzzin, $^{2}$
  T.~H.~Reiprich,$^1$
  M.~Schirmer,$^{10}$
  \newauthor
  P.~Schneider,$^1$
  A.~von der Linden$^{3,11,12}$ 
  \\$^1$Argelander-Institut f\"{u}r Astronomie, Rheinische Friedrich-Wilhelms-Universit\"at Bonn, Auf dem H\"{u}gel 71, D-53121 Bonn, Germany 
  \\$^2$Leiden Observatory, Leiden University, PO Box 9513, 2300 RA, Leiden, The Netherlands
  \\$^3$Kavli Institute for Particle Astrophysics and Cosmology, Stanford University, 452 Lomita Mall, Stanford, CA 94305-4085, USA.
  \\$^4$Steward Observatory, University of Arizona, Tuscon, AZ 85121, USA
  \\$^5$Jet Propulsion Laboratory, California Institute of Technology, 4800 Oak Grove Dr., Pasadena, CA 91109, USA
  \\$^6$Center for Cosmology and Astro-Particle Physics, The Ohio State University, 191 W. Woodruff Ave, Columbus, 43210 OH, USA
  \\$^7$Eureka Scientific, 2452 Delmer Street, Suite 100, Oakland, CA 94602-3017, USA
  \\$^8$Kavli Institute for Cosmological Physics, University of Chicago, 5640 South Ellis Avenue, Chicago, IL 60637, USA
  \\$^9$SLAC National Accelerator Laboratory, 2575 Sand Hill Road, Menlo Park, CA 94025, USA.
  \\$^{10}$Gemini Observatory, Casilla 603, La Serena, Chile
  \\$^{11}$Dark Cosmology Centre, Niels Bohr Institute, University of Copenhagen Juliane Maries Vej 30, 2100 Copenhagen {\O}, Denmark
  \\$^{12}$Department of Physics, Stanford University, 382 Via Pueblo Mall, Stanford, CA 94305-4060, USA
  }
\date{Submitted \today}

\def\LaTeX{L\kern-.36em\raise.3ex\hbox{a}\kern-.15em
    T\kern-.1667em\lower.7ex\hbox{E}\kern-.125emX}

\begin{document}

\label{firstpage}

\maketitle

\begin{abstract}
Finding a sample of the most massive clusters with redshifts $z>0.6$ can provide an interesting 
consistency check of the $\Lambda$ cold dark matter ($\Lambda$CDM) model. Here we present results from our search 
for clusters with 
$0.6\lesssim z\lesssim1.0$ where the initial candidates were selected by cross-correlating the RASS faint and bright 
source catalogues with red galaxies from the Sloan Digital Sky Survey DR8. Our survey thus covers 
$\approx10,000\,\rm{deg^2}$, much larger 
than previous studies \changes{of this \changestwo{kind}}. Deeper follow-up observations in three bands using the William 
\changes{Herschel} Telescope and the
Large Binocular Telescope were performed to confirm the candidates, 
resulting in a sample of 44 clusters for which we present richnesses and red sequence redshifts, as well 
as spectroscopic redshifts for a subset. At least two of the clusters in our sample are comparable in
richness to \mbox{RCS2-$J$232727.7$-$020437}, one of the richest systems discovered to date. We also 
obtained new observations with the Combined Array for Research 
in Millimeter Astronomy for a subsample of 21 clusters. 
For 11 of those we detect the Sunyaev-Zel'dovich effect signature. 
The Sunyaev-Zel'dovich signal allows us to estimate 
$M_{200}$ and check for tension with the cosmological standard model. We find no tension between our cluster 
masses and the $\Lambda$CDM model. 
\end{abstract}


\section{Introduction}
Clusters of galaxies, especially at high redshift, are important tools to study our Universe. 
Years before the discovery of dark energy in the late 20th century cluster studies  
already pointed towards an $\Omega_{\rm{m}}$ much smaller than unity (e.g.  \citealt{white}; \citealt{bahcall2}).
Furthermore, one can measure the total number of clusters per mass bin and compare it to theoretical predictions. 
\changes{In order to conduct such a cosmological analysis of a sample of galaxy clusters one first has to find them. 
Galaxy cluster detection is possible in many different ways depending on the wavelength. 
Since the intra-cluster medium (ICM) emits in the X-ray one can use X-ray surveys to detect clusters. This has been done 
many times using different X-ray observatories. For example using the \textit{ROSAT} satellite 
(e.g. XBAC: \citealt{xbac}; BCS: \citealt{bcs}; \changesthree{MACS: \citealt{macs}; }
HIFLUGCS: \citealt{reiprich}; 400D Cluster Survey: \citealt{burenin}) or 
the \textit{XMM Newton} satellite (e.g. XCS: \citealt{xcs}, \changesthree{\citealt{xcs_new}}; 
XMM LSS: \citealt{xmm_lss1}, \citealt{xmm_lss2}; REXCES\changestwo{S}: \citealt{rexcess}). 
Using the X-ray emission of the ICM one can measure the temperature of the gas, which probes the 
full gravitational potential of the cluster. Consequently, the X-ray properties of clusters correlate 
well with mass (e.g. \changestwo{\citealt{mahdavi}}). Once redshift, mass, and the selection function are known 
the samples can be used for constraining cosmological parameters (e.g. \citealt{viki}; \citealt{mantz}). 

Also, cosmic microwave background (CMB) photons experience inverse Compton scattering due to the electrons in the ICM and 
thus the CMB spectrum changes. Depending on \changestwo{the} frequency one will either observe a
decrease in photons or an increase. This \changestwo{is} known as the Sunyaev-Zel'dovich effect 
(SZE; \citealt{sz1}; \citealt{sz2}). 
The SZE is also being used as another way to find galaxy clusters
for example by the South Pole Telescope (SPT, e.g. \citealt{bleem}), the Atacama Cosmology Telescope 
(ACT, e.g. \citealt{hasselfield}) or 
the \textit{Planck} satellite \citep{planck_sz_cut}. \changesthree{T}he SZE probes 
the integrated pressure of the ICM\changestwo{, which} probes the gravitational potential and has also been found 
to correlate well with mass (e.g. \citealt{bonamente}). 
\changesthree{SZ-selected} samples have been used for cosmological parameter constraints 
(e.g. \citealt{benson}; \citealt{sievers}; \citealt{planck_sz_cosmo}). 

Galaxy cluster detection in the optical works somewhat \changestwo{differently}. Most cluster finding algorithms look for 
overdensities in the galaxy distribution. Nowadays, this is usually combined with magnitude information or 
photometric redshifts (e.g. \citealt{postman}; \citealt{3dmatched}). Similar to photometric redshifts one can 
also use colour information and an intrinsic property of clusters, the cluster red sequence. 
This red sequence can be observed as a region in the colour-magnitude diagram, where red galaxies of the same cluster
align along a line of almost constant colour \citep{red_sequence}. 
\changestwo{This is due to the redshift dependent shift of the $4000\,\rm{\AA}$-break through the filter bands in use, which 
is why the location of the red sequence in colour-magnitude space can be used as an estimator for the cluster redshift.}

The red sequence method has also been used for cluster detection for example by the 
Red Cluster Sequence Surveys 1 and 2 (\citealt{rcs}; \citealt{rcs2}), by the MaxBCG programme \citep{maxbcg}, or 
redMaPPer \citep{redmapper1}. 
\changestwo{\changesthree{Besides giving an estimate for the cluster redshift o}ptical surveys 
can \changesthree{also} provide estimates of ``cluster richness'',}
which is the number of 
cluster galaxies within a certain radius and brighter than some characteristic magnitude. 
\changestwo{Several cluster surveys have been generated around various richness measures
(e.g. \citealt{maxbcg}; 
\citealt{high}; \citealt{redmapper1}) and it has been shown to correlate with mass 
(e.g. \citealt{planck_sz_rich}, \citealt{act_rich}) although this relation appears to have large 
intrinsic scatter \citep{angulo}. }

Usually, the methods of cluster detection \changestwo{that} do not make use of optical observations require some kind of 
confirmation from a different wavelength regime. 
This can be overcome by cross-correlating data from two different regimes. This has been \changestwo{done} using optical 
and X-ray data by for example the Massive Cluster Survey (MACS, \citealt{macs}), the RASS-SDSS Galaxy Cluster Survey
\citep{popesso}, or the extended MACS (eMACS, \citealt{emacs}). Also optical and infrared data have been combined 
by the Massive Distant Cluster\changestwo{s} of Wise \changestwo{S}urvey (\citealt{brodwin}). 

\changestwo{
The most extreme clusters in mass ($M_{200}\geq5\times10^{14}M_{\odot}$) can be used for a cosmological test other than cluster 
counting. Given a cosmological model one can compute the allowed masses of galaxy clusters as a function of 
redshift (\citealt{haiman}; \citealt{weller}). This probes the extreme end of the mass function. 
In order to systematically search for the most massive clusters in our Universe a deep and wide area survey 
that probes large volumes needs to be carried out. Until recently, mostly samples consisting of only a few 
clusters that \changesthree{were} discovered in small surveys were tested for consistency with the $\Lambda$CDM model. 
For example\changesthree{,} \citet{broadhurst} used \changesthree{mass estimates based on}
strong lensing arcs of four galaxy clusters, whereas \citet{jee} used 
weak gravitational lensing \changesthree{masses} of 22 clusters. 
In \citet{mortonson} two clusters are tested and the authors provide 
a fitting formula for exclusion curves, which was shown to be too strict by \citet{hotchkiss}. 
In contrast to testing single cluster masses for consistency with the standard cosmological model one can also
use extreme number statistics and test a whole sample of clusters (\citealt{big_rad}; \citealt{order_stats}). 
So far only \citet{jee} find significant deviations from $\Lambda$CDM, using the exclusion curves from \citet{mortonson}. 
Considering the findings of \citet{hotchkiss} this tension has \changesthree{likely} been resolved. 

In the last years more large volume surveys were conducted. Especially the \textit{Planck} satellite 
has been shown to find massive galaxy clusters at redshifts greater than $z=0.5$ \citep{planck_sz_old} spread over 
the whole sky. 
This is complementary to the samples found by the SPT \citep{bleem} and ACT \citep{hasselfield}, 
which \changesthree{originate from} a smaller area and consist of \changesthree{typically slightly} 
less massive but higher redshift clusters. 

This work is meant to be a continuation of the still ongoing search for massive galaxy clusters at high redshift. 
By cross-correlating \changesthree{the positions of red galaxies in} the Sloan Digital Sky Survey (SDSS) and 
the faint and bright source catalogues of RASS, we create
a new sample of distant ($z>0.6$) and possibly massive cluster candidates, making use of the wide area of 
the SDSS Data Release 8. Because red galaxies are known to reside preferentially in clusters, this is a 
useful approach to identify massive clusters from the RASS catalogues \changesthree{which are} 
strongly contaminated with other X-ray sources 
(for example AGN or binary stars). 
Through follow-up observations using the William Herschel Telescope (WHT), the Large Binocular Telescope (LBT), and 
the Combined Array for Research in Millimeter Astronomy (CARMA), 
we then confirm or reject our candidates and check for consistency with $\Lambda$CDM. 
This study presents one of the first systematic searches for massive high-redshift galaxy clusters 
in the optical and X-ray regimes in a very large volume. 
}
Similar approaches to detect clusters have been used for eMACS \citep{emacs}, 
which also uses RASS data but for the optical part it makes use of 
deeper imaging data from the Pan-STARRS Medium Deep Survey, which is part of the 
Pan-STARRS project \citep{panstarrs}. Also, the aforementioned work by \citet{brodwin} searches for 
high redshift clusters in data from the \textit{Wide-Field Infrared Survey Explorer} (\textit{WISE}) satellite \citep{wise}. 
Instead of cross-correlating with optical data they use a non-detection in the SDSS as a\changesthree{n indication}
for a high-redshift cluster. 
}
 
\begin{figure}
 \centering
 \includegraphics[width=8.5cm,height=6cm,keepaspectratio=true]{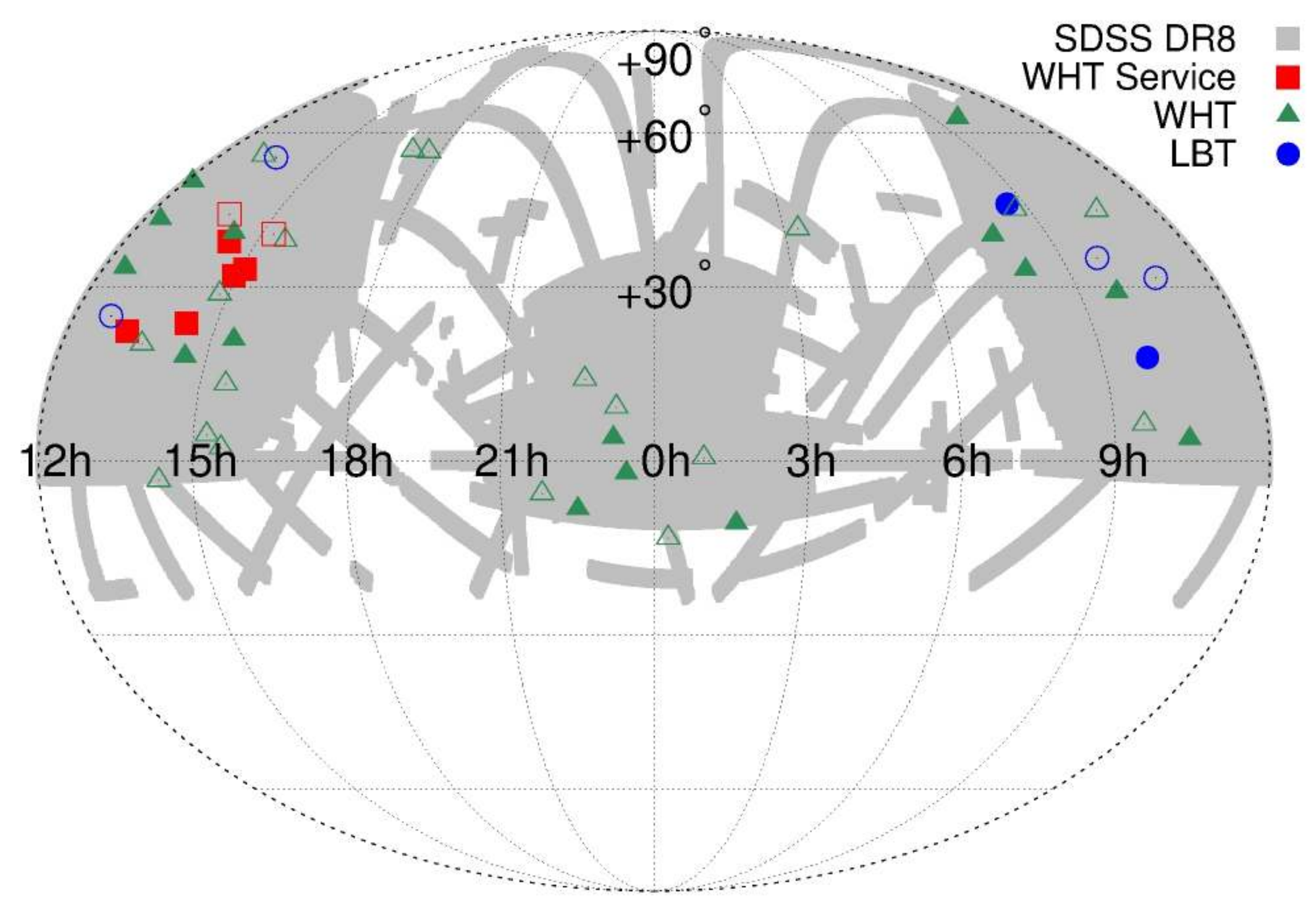}
 \caption{This plot shows the distribution of all clusters of our sample on the sky. Open symbols 
 indicate clusters with unknown spectroscopic redshift. Our search for clusters makes use of 
 about one quarter of the whole sky.}
 \label{fig:coords}
\end{figure}

One should note that we do not intend to use our sample for cosmological cluster abundance studies. 
By specifically following up the most extreme candidates we compromise a simple selection function. 
\changes{Nonetheless, it is one of the 
largest samples of very X-ray-luminous high-redshift galaxy clusters in the Northern hemisphere 
\changesthree{making it} complementary to the cluster samples found by 
\textit{Planck}, SPT, and ACT.} 
The distribution of all clusters in our sample on the sky is plotted in Fig. \ref{fig:coords}. 

In Section \ref{sec:sample}, we first describe how we define our cluster sample. We then explain the data 
from follow-up observations and the instruments which were used for those campaigns in Section \ref{sec:data}.  
This is followed by a detailed description  
about the red sequence and richness analysis and their interpretation in Section \ref{sec:analysis}. We describe 
the SZ data analysis in Section \ref{sec:sz}. In Section \ref{sec:outliers} we discuss possible tensions of our 
cluster sample with
$\Lambda$CDM and in Section \ref{sec:outstanding} properties of some individual clusters. This is followed by our 
conclusion. Images showing postage stamps of all 47 clusters, including three previously discovered objects, as well 
as SZ-maps from CARMA and \textit{Planck} data can be found in the appendix. 

As our fiducial cosmology we use $H_{0}=70\, \rm{km/Mpc/s}$, $h=0.7$, $\Omega_{\Lambda}=0.7$ and $\Omega_{\rm{m}}=0.3$. 
The exclusion plots in Section \ref{sec:outliers} were created assuming $\sigma_{8}=0.83$ as has been done 
in \citet{mortonson}. 
We define $r_{500}$ ($r_{200}$) as the radius, where the density of the galaxy cluster is 500 (200) 
times the critical density of the universe.

\section{Preselection of cluster candidates}
\label{sec:sample}
To find some of the most massive clusters at redshifts \mbox{$0.6\lesssim
z \lesssim1.0$}, we use the combined bright and faint source catalogues 
of RASS (\citealt{rass_bright}\footnote{\url{http://www.xray.mpe.mpg.de/rosat/survey/rass-bsc/}}; 
\citealt{rosat}\footnote{\url{http://www.xray.mpe.mpg.de/rosat/survey/rass-fsc/}}), which is an 
X-ray all sky survey in the $0.1-2.4$ keV range carried out with the \textit{ROSAT} satellite. 
This combined catalogue contains $125,000$ entries with typical positional uncertainties of $20\arcsec$. 
Most of these objects are not galaxy clusters but rather AGN or X-ray binaries.
Hence, to identify distant galaxy clusters, more information is needed. For that we 
combine the X-ray data with imaging data from the SDSS \citep{sloan}, where
we used Data Release 8 \citep{dr8}. 
By cross-correlating the RASS object positions with the position of
SDSS galaxies for which the SDSS photometry suggests that they likely match
the targeted redshift range,
we are able to efficiently preselect candidates for galaxy clusters.
Here we generally use a $50\arcsec$ matching radius, which should account for the
positional uncertainty in RASS and for the fact that galaxies scatter around the cluster centre. 
\changes{Note that we did not employ a radius in projected physical separation given the photometric redshift 
\changesthree{uncertainties} and the small change in projected radius of only 
about $50\,\rm{kpc}$ between $z=0.6$ and $z=0.9$.}
Photometric redshifts are taken from the \texttt{Photoz}-table in the SDSS archive. We then employ
two different SDSS galaxy selection schemes:
In the first scheme we select all SDSS galaxies with a photometric redshift
\mbox{$z>0.6$} and \mbox{$i<20.5$}. This yields 1149 matches of RASS
sources with two or more SDSS DR8 galaxies, mostly at
\mbox{$0.6\lesssim z \lesssim 0.8$}.
At higher redshifts we expect that possibly only a single cluster galaxy
(the BCG) is detected in SDSS. We select candidates for such galaxies
photometrically from SDSS with colour cuts \mbox{$r-i>0.5$},
\mbox{$i-z>0.8$}, and
\mbox{$17<i<21$} (compare e.g. \citealt{high}). 
While requiring a match of at least one of these galaxies in the
SDSS DR8 with the RASS sources and adding these cases to our preselected sample 
we find 1395 candidates in total. 

In the next step all candidates are visually inspected using SDSS postage stamps and graded.
Here we immediately drop obvious chance alignments of background
galaxies e.g. with bright foreground stars, spectroscopically classified
QSOs, or low-$z$ galaxy groups, which most likely dominate the
X-ray flux. 
In addition, we drop sparse galaxy groups/clusters, where the SDSS colours
suggest \mbox{$z\sim 0.6-0.7$}. At these redshifts we would still expect to
detect numerous cluster galaxies in SDSS if these were massive clusters. Hence, these sparse  
groups/clusters likely have an
X-ray flux boosted by an AGN and are not of interest for our study.
The remaining candidates are graded in preparation for further 
follow-up observations (Section\thinspace\ref{sec:data}), where we prioritize the 
richest systems as well as good candidates for the highest-redshift
clusters (\mbox{$z\gtrsim 0.8$}) in our sample. 
\changesthree{We attempted optical follow-up observations for a total of 80 candidates. 
From these 48 have data of sufficient quality in the three filters $r,i,z$, constituting the 
sample we analyse in this paper. This includes all of the top-graded candidates.
For eight of the remaining candidates, single band observations were sufficient to identify them as 
false positive. The remaining 24 candidates, which were all of lower or medium priority, were 
dropped from the current analysis, as they do not have observations of sufficient quality in all 
three bands. This was due to observations attempted under poor conditions, guiding errors, or limited 
target visibility. Within the allocated time these observations could not be completed or repeated, 
but we ensured to complete the observations for all of the highly-graded candidates. }

With our automated pre-selection we also `rediscovered' the known massive
clusters 
MACS$J$0744.8+3927 (\mbox{$z=0.6976$}; \citealt{red_macs}),
MACS$J$2129.4$-$0741 (\mbox{$z=0.5889$}; \citealt{red_macs}),
and \mbox{RCS2-$J$232727.7$-$020437},
\citep[\mbox{$z=0.705$;}][]{2327},
providing a confirmation of our algorithm and a reference sample of massive
clusters in the targeted redshift range.

\section{Follow-up observations}
\label{sec:data}

\subsection{Optical images}
\subsubsection{William Herschel Telescope}
\label{sec:data:imaging:wht}

The majority of our optical follow-up observations were taken with the Auxiliary-port CAMera (ACAM) \citep{acam} on the 
4.2-m William Herschel Telescope on the island of La Palma in Spain.
ACAM is a red-optimized one chip camera with $2148\times2500$ pixels which has an unvignetted circular field of 
view of about $8\arcmin$ in diameter and a pixel scale of $0\farcs25$. 
 
Our WHT data were taken in service mode (August 2010 and August 2013, PIs Schrabback and 
Buddendiek, respectively), and in visitor mode (four nights each in August 2011 and March 2012, PI Schrabback). 
We obtained imaging in $r$, $i$, and
$z$ filters, which bracket the $4000\,\rm{\AA}$-break in the redshift range of interest. 
The service observations in 2010 were carried out with the RGOZ2 filter 
($\lambda_{\rm{central}}=8748\,\rm{\AA}$) as the
SDSS $z$-band was not yet available. Therefore we need to create different red 
sequence models for those images later on. 
Our total exposure time per cluster candidate per filter varies between 360 and 1800
seconds, this choice primarily depends on observing conditions and the roughly estimated
cluster redshift.
For some of the candidates for the  highest-redshift clusters in the sample
-- which typically were the most 
uncertain
candidates with only a single noisy BCG candidate -- we stopped observing
after taking data in a single filter ($i$ or $z$) if these data clearly
showed that this was a spurious match (e.g. a faint red star misclassified
as galaxy in SDSS).  
In total we obtained 3-band imaging for \changes{42} cluster candidates with
ACAM, plus 3 previously known clusters with  spectroscopic redshifts
which were included as reference objects for the generation of the red
sequence model (see Table 1).

\subsubsection{Large Binocular Telescope}
We observed nine cluster candidates using the \mbox{$2\times 8.4$-m} Large
Binocular Telescope in Arizona during observations in October and December 2010, as
well as February and April 2011 (PI: Eifler). Two of these candidates were also observed with the WHT. 
Here we employed the $r$-, $i$- and $z$-filters, which are similar to the WHT 
filters used. The instruments
used were LBC\_RED ($i$- and $z$-band) and LBC\_BLUE ($r$-band) \citep{lbc}. Those cameras have 
four $2048\times4608$ pixel 
chips each, a pixel scale of $0\farcs23$ and a field of view of about $24\times25$ $\rmn{arcmin}^{2}$. 
A single chip covers roughly $17\times8$ $\rmn{arcmin}^{2}$. 

Total exposure times per filter for the LBT data are between 360 and 720 seconds, depending on the object. 
Single exposures were integrated for $180$ seconds regardless of the filter in use. 

\subsection{Spectroscopic observations}
We obtained long-slit spectroscopic data for 14 clusters with ACAM 
during the visitor mode WHT runs listed in
Sect.\thinspace\ref{sec:data:imaging:wht}, plus one cluster as part of a WHT
service program in June 2014 (PI: Buddendiek). Targets were selected for the spectroscopic observations 
either if they appeared to be very rich, at very high redshift or if they seemed relaxed due to a single very bright 
BCG. 
Integration times varied between $600\,\rm{s}$ and $1100\,\rm{s}$ per
exposure, which results in total integration times between $1800\,\rm{s}$ and $3300\,\rm{s}$ per target. 
In all cases we employed the V400 grating and the G495 filter, which provides 
a wavelength range from $4950\,\rm{\AA}$ to $9500\,\rm{\AA}$ and $3.3\,\rm{\AA}/\rm{pixel}$. The slit width is
$1\farcs0$, corresponding to a resolution of $R=570$ at a wavelength of $\lambda=7500\,\rm{\AA}$. 
For three clusters the spectra are 
too noisy and no redshift could be estimated. We generally placed the slit on
top of the BCG
and if possible oriented it such that 
other cluster members were visible through the slit as well. 

\subsection{Data reduction and calibration}
\label{sec:reduction}

The WHT and LBT data are reduced using the GUI version of the 
THELI\footnote{\url{http://www.astro.uni-bonn.de/~theli/index.html}}
pipeline (\citealt{theli}; \citealt{theli2}). We apply bias subtraction, flat-field 
correction, and superflat field correction. Exposures are co-added and later convolved with 
a Gaussian kernel to have approximately the same resolution in all bands for photometric measurements. 

We calibrate the photometry by fitting the function
\begin{equation}
 \rm{mag}_{\rm{SDSS}}-\rm{mag}_{\rm{m}}=\rm{C}_{\rm{SDSS}} \cdot \rm{CT} + \rm{ZP}
\end{equation}
to field stars. $\rm{mag}_{\rm{m}}$ is the measured magnitude, 
$\rm{mag}_{\rm{SDSS}}$ the corresponding
SDSS magnitude, CT the colour term and ZP the magnitude zero-point. $\rm{C}_{\rm{SDSS}}$ is the SDSS colour we
use for calibration, either $r-i$ ($r$- and $i$-band calibration) or $r-z$ ($z$-band calibration). 
After correcting magnitudes 
with the zero-points we do not apply a colour correction but work in the instrumental system instead. 
Every single field is corrected independently. 
The data reduction for WHT and LBT data is performed in the same way. 

In order to determine the limiting magnitude of a co-added image we use 
\begin{equation}
\label{eq:lim_mag}
 m_{\rm{lim}} = \rm{ZP} - 2.5 \log \left(5\sqrt{N_{\rm{pix}}}\sigma_{\rm{sky}}\right), 
\end{equation}
where $N_{\rm{pix}}$ is the number of pixels within a circle with a radius of $2\farcs0$ and $\sigma_{\rm{sky}}$
is the variation of the sky background noise (see \citealt{mag_lim}). This gives the $5\sigma$ detection limit. 
We find the mean limiting magnitudes of the WHT images to be $r_{\rm{lim}}=23.81\,\rm{mag}$, 
$i_{\rm{lim}}=23.42\,\rm{mag}$ and \mbox{$z_{\rm{lim}}=22.64\,\rm{mag}$}.
We also measure the seeing as the \texttt{FWHM} and find the median seeing $\mathtt{FWHM}_{r}=0\farcs95$, 
$\mathtt{FWHM}_{i}=0\farcs82$ and $\mathtt{FWHM}_{z}=0\farcs82$.
For the LBT data we find $r_{\rm{lim}}=24.52\,\rm{mag}$, 
$i_{\rm{lim}}=24.95\,\rm{mag}$, $z_{\rm{lim}}=23.63\,\rm{mag}$ and $\mathtt{FWHM}_{r}=0\farcs77$, 
$\mathtt{FWHM}_{i}=0\farcs92$, $\mathtt{FWHM}_{z}=0\farcs77$. 

The spectra are also bias subtracted, flat fielded and then extracted. For the further reduction we use 
\textsc{IRAF} \citep{iraf2}. We extract the spectra using the 
task \texttt{apall}. Furthermore, wavelength and flux calibration are performed with the tasks
\texttt{identify}, \texttt{dispcor} and \texttt{calibrate} using skylines and standard star observations. 

\subsection{Sunyaev-Zel'dovich data}
To obtain cluster mass estimates, we targeted a sub-sample of 21
targets with the Combined Array for Research in Millimeter-wave
Astronomy to measure the SZE
signal, which has been found to correlate with mass with small
intrinsic scatter, both from simulations (e.g. \changes{\citealt{dasilva}}; \citealt{motl}; \citealt{stanek}) 
and observations (\changes{e.g. \citealt{bonamente}; \citealt{planck_sz_obs1}; \citealt{marrone}; 
\citealt{planck_sz_obs2}}). 

The SZE data for 20 of those clusters were obtained using the eight 3.5-m telescopes of CARMA
in the SH and SL configurations.  For these configurations, six
telescopes are grouped in a compact central array and two on outlying
pads. The long baselines resolve out the cluster signal and yield
uncontaminated measurements of point sources, which can then be
subtracted from the short baseline data. We used the CARMA wideband
correlator with $8\,\rm{GHz}$ of correlation bandwidth.  Observations were
carried out in the \mbox{$30\,\rm{GHz}$} band and integration times were planned to
be $8\,\rm{h}$ for each cluster.  Due to various
reasons the $8\,\rm{h}$ were not always reached. The exact integration
times can be found in Table \ref{tab:carma}. The CARMA programme numbers
are \texttt{c0734}, \texttt{c0734Z} (both PI: Schrabback) and
\texttt{c0934} (PI: Plagge). Those targets were selected because they appeared 
to be the richest or most distant objects in the sample. 
Additionally, we also have been granted director's discretionary time for the target 
ClG-$J$122208.6$+$422924 (\texttt{cx389}, PI: Buddendiek). 
\changesthree{This data set was recorded using an antenna configuration
different from the SL and SH configurations. All 3.5m-antennas were grouped}
in a compact array and the 6-m and 10-m antennas are used for long baselines. 

\changes{The first 20 targets were selected after an initial optical analysis because they 
appeared to be either the richest, the most X-ray luminous or the highest redshift ones. 
One should note that at that time the optical campaign was
not complete yet. The last of the 21 targets was selected after the optical analysis had been completed and it 
had a measured spectroscopic redshift greater than 1, which is the highest in the whole sample. }

\section{Optical Data Analysis}
\label{sec:analysis}

\subsection{Spectroscopic redshifts}
After extracting the spectra we use the \textsc{IRAF} task \texttt{fxcor} \citep{fxcor} in order to cross-correlate them
with the absorption line template spectrum \texttt{fabtemp97} and the emission line template spectrum 
\texttt{femtemp97}.
This yields the redshift estimates. In order to find the uncertainty \texttt{fxcor} fits a Gaussian to the
correlation peak and we then take the half-width at half-maximum as the redshift error. 
Visually identified lines and features can be found in Table \ref{tab:spec_z}. 

The spectra are mainly low S/N spectra due to very faint targets. The redshifts are mostly estimated using
absorption features like the Ca K+H doublet, thus the errors for the redshifts are comparably high 
($\approx0.5$ per cent). Individual errors can be found in Table \ref{tab:spec_z}. 

In our analysis we also include the already known redshifts of twelve galaxy clusters. Those were taken either 
from the SDSS Data Release 10 \citep{dr10} or from other independent discoveries. In one of those cases (ClG-$J$131339.7$+$221151) a 
spectrum from the SDSS was available but no reliable redshift has been estimated 
(\mbox{$z_{\rm{SDSS}}=1.000\pm3.359$}); 
we downloaded the already reduced and extracted spectrum and estimate the redshift ourselves. 
All redshifts used in this study are listed in Table \ref{tab:spec_z}, which also includes additional information.
\begin{table*}
\begin{minipage}{15cm}
\caption{The spectroscopic sub-sample. Spectroscopic redshifts are either measured from our data, taken 
          from independent discoveries or from the SDSS DR 10. If $z_{\rm{spec}}$ was measured, the spectroscopic
          features which were identified by visually inspecting the spectra are listed. 
          For ClG-$J$131339.7$+$221151 we downloaded one spectrum from 
          the SDSS data base and determined the redshift ourselves, because the estimate taken from SDSS 
          proved not to be trustworthy ($z_{\rm{SDSS}}=1.000\pm3.359$). }
\begin{center}
\begin{tabular}{|l|c|c|c|c|}
\hline \hline
Object & \multicolumn{1}{l|}{Redshift} & Lines & \# Spectra & Ref. \\ \hline
ClG-$J$013710.4$-$103423 & 0.662$\pm$0.002 & Ca H+K, 4000$\rm{\AA}$ & 1 & - \\ \hline
ClG-$J$031924.2$+$404055 & 0.680$\pm$0.003 & Ca H+K, 4000$\rm{\AA}$ & 1 & - \\ \hline
MACSJ0744.8$+$3927\footnote{\label{note}These clusters were known before and are only included
in the sample for calibration reasons. } & 0.698 & - & \multicolumn{1}{c|}{-} & \citet{red_macs} \\ \hline
ClG-$J$080434.9$+$330509 & 0.553 & - & \multicolumn{1}{c|}{1} & SDSS \\ \hline
ClG-$J$083415.3$+$452418 & 0.666 & - & 1 & SDSS \\ \hline
ClG-$J$094700.0$+$631905 & 0.710 & - & \multicolumn{1}{c|}{1} & SDSS \\ \hline
ClG-$J$094811.6$+$290709 & 0.778$\pm$0.002 & Ca H+K, 4000$\rm{\AA}$ & 1 & - \\ \hline
ClG-$J$095416.5$+$173808 & 0.828 & - & \multicolumn{1}{c|}{-} & \citet{0954} \\ \hline
ClG-$J$102714.5$+$034500 & 0.749$\pm$0.003 & Ca H+K, 4000$\rm{\AA}$ & 1 & - \\ \hline
ClG-$J$120958.9$+$495352 & 0.902$\pm$0.001 & [OII], Ca H+K & 1 & - \\ \hline
ClG-$J$122208.6$+$422924 & 1.069$\pm$0.003 & Ca H+K, 4000$\rm{\AA}$ & 2 & - \\ \hline
Cl$J$1226.9$+$3332\textsuperscript{\ref{note}} & 0.892 & - & \multicolumn{1}{c|}{-} & \citet{red_clj} \\ \hline
ClG-$J$131339.7$+$221151 & 0.737$\pm$0.002 & Ca H+K, 4000$\rm{\AA}$ & 1 & SDSS \\ \hline
ClG-$J$142040.3$+$395509 & 0.607 & - & \multicolumn{1}{c|}{-} & \citet{red} \\ \hline
ClG-$J$142138.3$+$382118 & 0.762 & - & \multicolumn{1}{c|}{1} & SDSS \\ \hline
ClG-$J$142227.4$+$233739 & 0.726 & - & \multicolumn{1}{c|}{1} & SDSS \\ \hline
ClG-$J$143411.9$+$175039 & 0.744$\pm$0.003 & Ca H+K, 4000$\rm{\AA}$ & 1 & - \\ \hline
ClG-$J$145508.4$+$320028 & 0.654 & - & \multicolumn{1}{c|}{1} & SDSS \\ \hline
ClG-$J$150532.2$+$331249 & 0.758 & - & 1 & SDSS \\ \hline
ClG-$J$152741.9$+$204443 & 0.693$\pm$0.002 & Ca H+K, 4000$\rm{\AA}$ & 1 & - \\ \hline
ClG-$J$223007.6$-$080949 & 0.623$\pm$0.003 & Ca H+K, 4000$\rm{\AA}$ & 1 & - \\ \hline
ClG-$J$231215.6$+$035307 & 0.648$\pm$0.003 & [OII], Ca H+K, 4000$\rm{\AA}$ & 4 & - \\ \hline
RCS2-$J$232727.7$-$020437\textsuperscript{\ref{note}} & 0.705 & - & \multicolumn{1}{c|}{-} & \citet{2327} \\ \hline
\end{tabular}
\end{center}
\label{tab:spec_z}
\end{minipage}
\end{table*}

\subsection{Red sequence finding and redshift estimation}
We derive empirical red sequence models in $r-i$, $i-z$ and $r-z$ using 12 clusters from the WHT sample 
with known 
spectroscopic redshifts. For this we use the colour-magnitude diagram of galaxies within the inner $50\arcsec$ around
the BCG. \changes{Again, we employ a constant angular radius and not a physical one given the \changestwo{small} 
change in the angular diameter distance between $z=0.6$ and $z=0.9$. }

\begin{figure}
\centering
 \includegraphics[width=9cm,height=7cm,keepaspectratio=true]{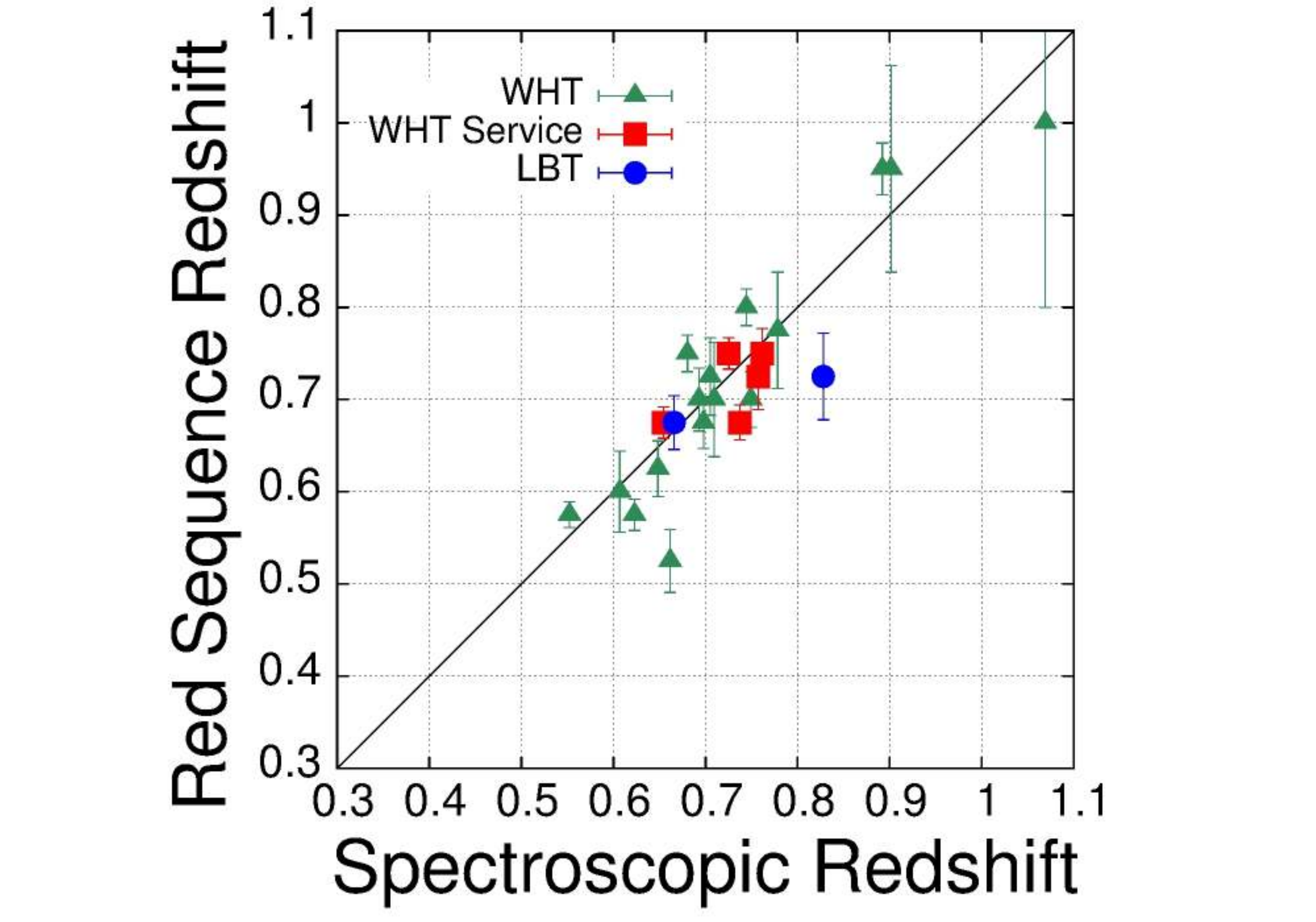}
\caption{Comparison of spectroscopic versus red sequence redshifts of galaxy clusters. 
         Error bars represent statistical errors and photometric errors, 
          which originate from the photometric calibration. 
          The black line shows the one on one relation. No systematic bias seems to be present.
          }
\label{fig:calib}
\end{figure}

\changes{\changesthree{Within} this radius} we fit a linear function of \changesthree{galaxy} colour versus magnitude as a 
red sequence yielding slope and offset. 
We then assume that red sequence slope and offset change linearly with redshift and thus fit both as a linear 
function of $z$. 
Using these fits we can derive an empirical red sequence model for every redshift 
in the range $0.5\lesssim z \lesssim0.9$. Additionally, we extrapolate these models to $z=0.4$ and $z=1.0$. 
\changes{We are aware that the red sequence slope and off-set do not in general vary linearly with
redshift. Nevertheless, this assumption provides a good approximation given the redshift range and 
filter choice.}
The models created can be used for both the WHT and the LBT sample, because their 
filter sets are fairly similar; for the service observations in 2010, we create models in the same way but 
using different clusters, due to the different filters used. \changes{The clusters used 
\changestwo{to create the models} for the WHT and LBT
samples spread almost evenly in the \changesthree{redshift range between} $z\approx0.55$ and $z\approx0.9$. For the models for 
the WHT service observations we only have redshifts available between $z\approx 0.6$ and $z\approx0.8$. 
Later on in this section we will find these models to be sufficient for our purposes (see Fig. \ref{fig:calib}). }

We create the galaxy catalogue with aperture photometry in dual image mode, using
the $i$-band as the detection image. Due to the homogenized PSF we suppress background noise and thus 
underestimate the photometric errors. To avoid this issue we run \texttt{SExtractor} \citep{bertin} 
again on the unconvolved 
images and use those magnitude errors. Nevertheless, we find that we still underestimate the photometric errors due to 
multiple reasons. For example, during the reduction we resample the images to a new pixel grid, which correlates 
the background noise. 
This has a similar effect as the PSF homogenization. We also use aperture photometry, which can lead to additional 
photometry errors, in case of a not completely homogeneous PSF in all three filters. In order to account for this, 
we take the photometric errors from \texttt{SExtractor} to be twice as large as the original value. 
\changes{A factor of $1.3$ is due to noise correlations, the remaining due to 
\changesthree{uncertainties arising from the limitations in the} PSF homogenization\changesthree{. 
This is performed} by assuming Gaussian PSFs and 
by quantifying the PSF using the flux radius, which is not a complete description of the PSF. 
In the end this results in a total correction factor of 2. }
Using the newly created models, we find the red sequence and the corresponding redshifts by taking the 
following steps, which are similar to the approach used in \citet{high}: 

First we identify the BCG in the colour image. 
We then use all galaxies, which are within a given radius $R$ around the BCG. 
Additionally, we only take galaxies with an $S/N$ larger than 
6 in the
$i$-band into account. 
Between redshifts 0.4 and 1.0, we proceed in steps of $\Delta z=0.025$ and use the 
corresponding red sequence model to look for galaxies in the catalogue which lie within a
certain error range in colour, $\Delta c$, from the red sequence lines in all three colours. 
Here, we also use galaxies even if they only fall within that range, when taking their magnitude errors into
account. 
Although we only use the inner parts around the cluster centre we are still affected by 
fore- and background galaxies, which are contaminating the colour-magnitude diagram. 
In order to avoid false detections through these galaxies, we determine and subtract an average 
red sequence background. Since the ACAM field of view is fairly small, we use about 100 apertures in the public 
CFHTLenS catalogue (\citealt{erben}; \citealt{hilde}), using the same cuts as for the actual galaxy catalogues 
in order to estimate the mean red sequence object density. 
After normalizing by the projected area and subtracting the background, we choose the redshift bin which contains the
most galaxies to be our red sequence redshift estimate. 
The error range $\Delta c$, and the aperture 
radius $R$ are free variables, which can be chosen arbitrarily. We explore the parameter space 
spanned by those two parameters, looking for the combination which recovers the known spectroscopic 
redshifts best. Although we vary the radius $R$ for each cluster, we find that the best choice for all the
WHT objects is $R=1\farcm25$ and $R=0\farcm76$ for all the LBT targets. 
While looking for the red sequence for every cluster candidate, we maximize the signal by varying $\Delta c$ 
in discrete steps between $0.01$ and $0.2$. In the end for each cluster we pick the value, which leads to the 
strongest signal. \changes{A typical value here is $\Delta c=0.08$}. 

We plot the estimated spectroscopic redshifts against their measured photometric counterparts for the best configuration
of $R$ and $\Delta c$.  
As can be seen in Fig. \ref{fig:calib}, no systematic bias is present, and on average the 
red sequence redshift estimates agree with the spectroscopic ones. Thus, we decide not to calibrate the estimates
further. 

The comparison with the spectroscopic sample shows that the models work fine as we find
$\sigma_{z}=0.037$, which we define as
\begin{equation}
 \sigma_{z} =\sqrt{\frac{1}{N}\sum \left( \frac{z_{\rm{spec}}-z_{\rm{phot}}}{1+z_{\rm{spec}}}\right)^{2}}, 
\end{equation}
where $N$ is the number of galaxy clusters with a known spectroscopic redshift and $z_{\rm{spec}}$ and $z_{\rm{phot}}$ 
is their corresponding spectroscopic or red sequence redshift. 

We also try building analytical models from \citet{bc03}, taking into account filter curves, quantum efficiency, and 
reflection curves of all optical elements inside the telescope, but we found that, especially at the 
low- and high-redshift regions in our sample, the redshift estimation failed completely. These models apparently 
do not match the observed galaxy distribution over the whole redshift range. Already 
\citet{photoz} showed that photometric redshift codes, which are tested on a suitable training sample, usually work 
best while using empirical models. In the end we decided to use the empirical models rather than the analytical ones. 
 
A colour image of a typical cluster, a background subtracted histogram of possible red sequence members,
the red sequence corresponding to the photo-z estimate and also the number counts (Section \ref{sec:rich}) 
can be seen in Fig. \ref{fig:2312_all}. \\
We estimate statistical errors from bootstrapping the whole galaxy catalogue and 
estimating the redshift several thousand times. To the standard deviation of the distribution, which is the statistical 
error, 
we quadratically add the magnitude zero-point error, which gives a fair estimate of the photometric error, and take this 
as the red sequence redshift uncertainty. We check if this is indeed a fair representation of the true uncertainty by
computing the standard deviation, $\Delta z$, of $z_{\rm{spec}}-z_{\rm{phot}}$ and comparing it with the mean 
redshift error $\langle \Delta z\rangle$. We find $\Delta z=0.048$ and $\langle \Delta z\rangle=0.044$. 
This means that on average $\Delta z$ is a good representation of the true redshift uncertainty. 

\subsubsection{Defining a detection}
After running our red sequence finder on the data of all 48 cluster candidates, which have three-band imaging, 
we define a detection using two criteria:
\begin{enumerate}
 \item The object shows a peak in the red sequence histogram (see Fig. \ref{fig:2312_all}, top-right panel). 
 \item In the three-colour image, we can visually find an over-density of galaxies, which have the same colour. 
\end{enumerate}
If both these criteria are true, we consider this a detection and continue the analysis. 
If only one or none are true, we stop the analysis after the red sequence finding and consider this a non-detection. 
From the 48 cluster candidates, we detect 44 according to these criteria. The three previously known clusters 
are detected as well.

\subsection{Richness estimates}
\label{sec:rich}
We define the richness $N_{\rm{gal}}$ to be the number of cluster galaxies within 0.5 Mpc around the BCG, 
which are brighter than
some characteristic magnitude of the cluster luminosity function. We will now describe the procedure to estimate 
$N_{\rm{gal}}$. 

\begin{figure*}
 \centering
 \includegraphics[width=18cm,height=12.6cm,keepaspectratio=true]{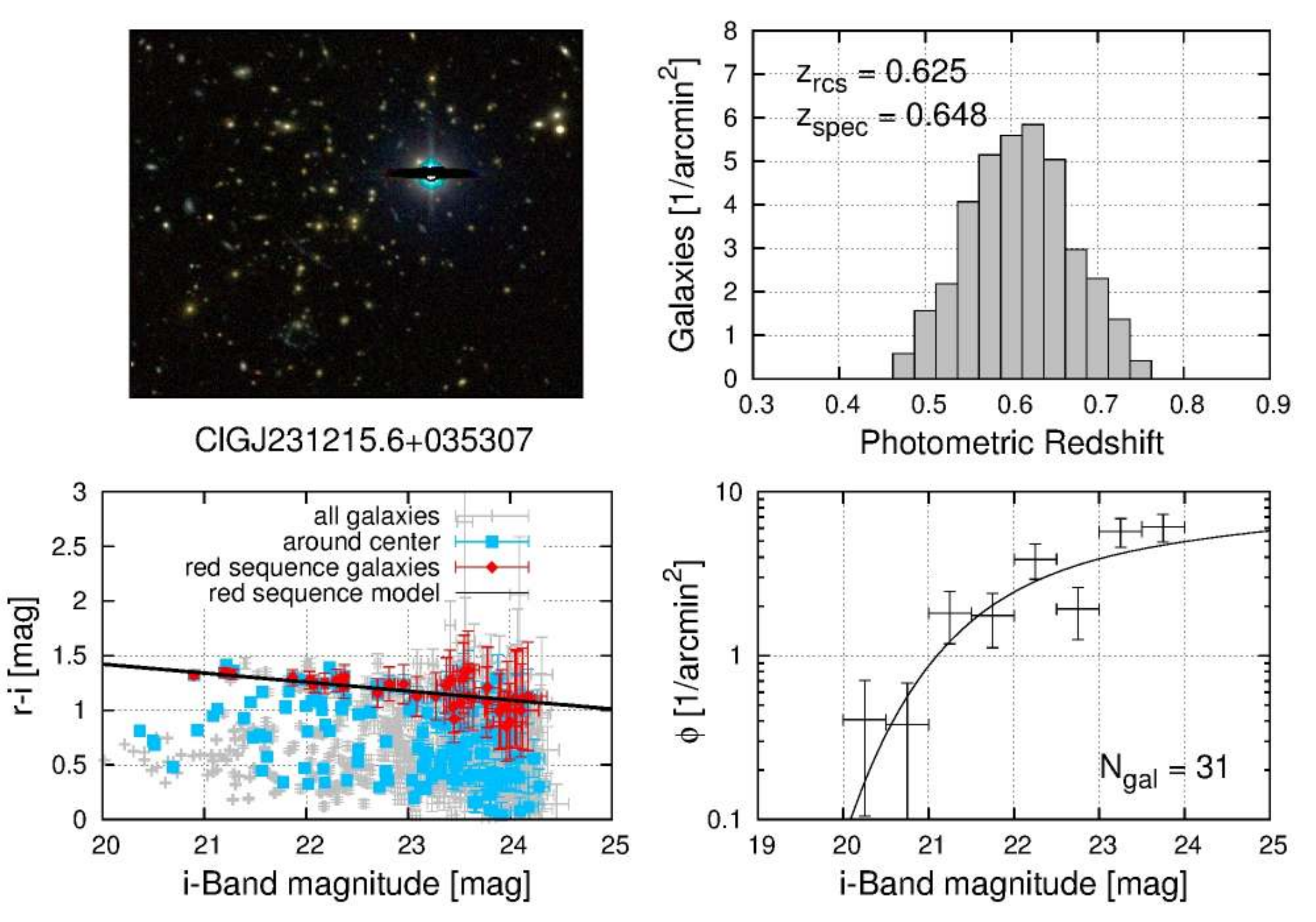}
 \caption{This figure shows the output of the red sequence analysis for one cluster, 
          ClG-$J$231215.6$+$035307. The top-left panel shows a colour image of the inner parts of the cluster. 
          In the top-right panel, we show the number of galaxies around the cluster centre, which coincide with the
          red sequence models as a function of redshift. Here the peak lies at $z=0.625$. 
          The bottom-left panel shows a colour-magnitude
          diagram. Grey points are all galaxies in the field, blue points are galaxies within $1\farcm25$ of the 
          centre and red points are red sequence galaxies. 
          The black line shows the red sequence for $z=0.625$. 
          Finally, the bottom right panel shows the $i$-band number counts of the cluster members, 
          shown in the figure to the left. The black line is the best Schechter function fit. 
          The fact that the number counts do not start to decrease at fainter 
          magnitudes suggests that we do not suffer from significant incompleteness issues. }
 \label{fig:2312_all}
\end{figure*}

Once the red sequence redshift was estimated, we created new catalogues with all galaxies which were 
detected as a red sequence member in all three colours at this redshift. 
For the aperture radius $r$, we now choose 0.5 Mpc. 
The galaxies are divided 
in magnitude bins of size 0.5 mag between 19th and 24th magnitude in the $i$-band and normalized to the area. 
Again, a background is estimated from CFHTLenS and subtracted. We then fit a Schechter function 
\citep{lum_func} normalized to projected area rather than volume to the data
\begin{eqnarray}
    \phi(m) \ \mathrm{d}m & = & 0.4 \ \ln 10 \ \phi^* 10^{ -0.4 ( m - m^* ) \cdot ( \alpha + 1)} \\
    & & \times \exp [ -10^{ -0.4 ( m - m^* ) } ] \ \mathrm{d}m \nonumber . 
\end{eqnarray}
\changes{For the fit we keep $\alpha$ fixed to $-1.1$, which has been shown to be robust for rich clusters 
(e.g. \citealt{paolillo}). 
Furthermore, we assume passive stellar evolution and use the stellar population synthesis models from 
\citet{bc03} with the Padova stellar evolution models \citep{padova} and the initial mass function by \citet{imf} to
fix $m^{*}$ for every redshift. In the end, we only fit the normalization $\phi^*$. }
Subsequently, we integrate the Schechter function 
up to $m^{*}+2$. 
After multiplying the result with the projected area this gives us our richness estimate, $N_{\rm{gal}}$. 
An example of such a measured function can be found in the bottom right panel of 
Fig. \ref{fig:2312_all}. 

We estimate statistical errors for the richness by bootstrapping the cluster member 
sample and repeating the whole estimation procedure several thousand times. We then quadratically add the Poissonian
error and take this as the total uncertainty in richness. 
For comparison, we also estimate the richness of a cluster by counting the red sequence galaxies that are brighter 
than $m^{*}+2$ and call this $N_{\rm{count}}$. Here we take the Poissonian error as the uncertainty. 
For the further analysis we use only the $N_{\rm{gal}}$ estimates, because we expect them to be more robust. 

Redshifts, richnesses and other properties as well as comments concerning the data and the analysis can be found 
in Table \ref{tab:rich_red}. 

\subsection{Discussion of the results from the optical data}
With our analysis we confirmed 44 galaxy clusters at redshifts between $0.5\lesssim z \lesssim1.0$. Additionally, we 
conducted the analysis for three previously known clusters in order to have a calibration sample. 
The cluster richnesses within 0.5 Mpc vary between 3 and 46. We summarize all measured quantities in 
Table \ref{tab:rich_red}. One column in this table lists problems that occurred during the
analysis. Those problems were poor observing conditions like high airmass, 
cloud coverage etc., which lead to considerable systematic uncertainties. 
Furthermore, the galaxy redshift distribution 
in the histograms like the one shown in Fig. \ref{fig:2312_all} does not always have a clear peak, sometimes 
it is bimodal. 
\changestwo{Additionally, the Schechter function fit can fail, which can for example be caused by a 
poor redshift estimate due to a faint cluster. An example for this is ClG-$J$094742.3$+$351742. 
From the fit we find $N_{\rm{gal}}=20\pm4$, which does not agree with the counted estimate of $N_{\rm{count}}=2\pm1$. 
\changesthree{Poor} data in one or more bands can also lead to \changesthree{poor} richness estimates. 
The $r$-band of ClG-$J$144847.4$+$284312
for example is much shallower than the rest of the data\changesthree{,} because it 
\changesthree{was} observed in bright time. 
Due to this we overestimate the background in this field, which leads to 
the low values in $N_{\rm{gal}}=3\pm2$ and $N_{\rm{count}}=3\pm2$. 
}

\begin{figure}
 \centering
 \includegraphics[width=8cm,height=5.6cm,keepaspectratio=true]{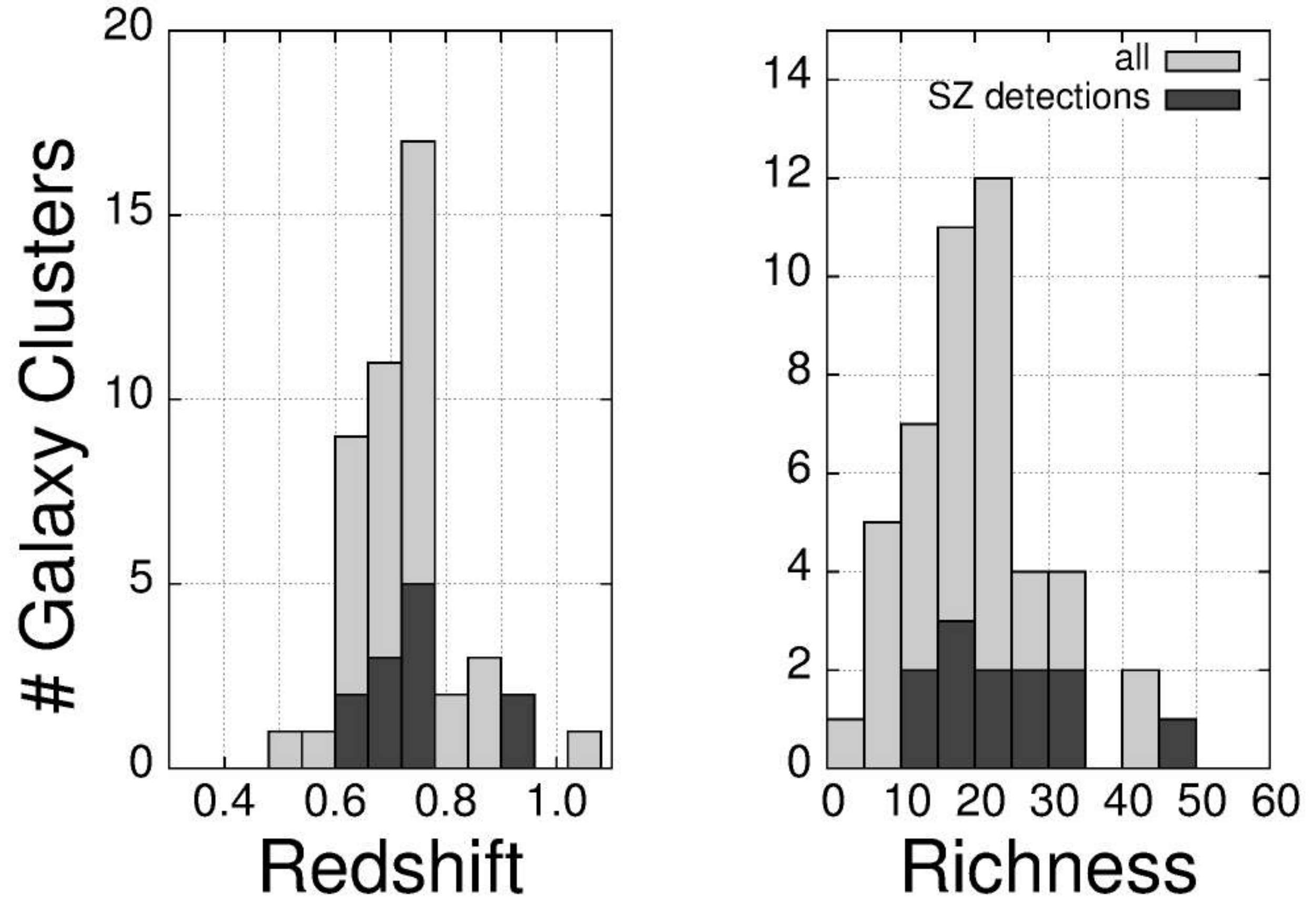}
 \caption{Redshift and richness distribution of all 44 galaxy clusters in our sample and of the three 
          previously known objects. 
          Wherever available we use spectroscopic redshifts. 
          Light grey bars show the whole sample, dark grey bars the SZ-detected clusters only.}
 \label{fig:distros}
\end{figure}

The redshift and richness distribution of our sample can be found in Fig. \ref{fig:distros}. The redshift 
distribution
peaks at \mbox{$z=0.75$}. We targeted a redshift range of $0.6\lesssim z\lesssim 1.0$ while cross-correlating RASS and SDSS. 
In this respect the left-hand panel of Fig. \ref{fig:distros} is a confirmation that our approach works indeed. 
The richness distribution shows a peak between 20 and 30 and then a decreasing trend towards higher richness. 
The most interesting objects are those at the high richness tail at $N_{\rm{gal}}>30$. 
Nevertheless, all objects in this sample seem to be rare X-ray luminous high-redshift galaxy clusters, which makes them 
interesting objects for further research. 

\begin{figure*}
 \centering
 \includegraphics[width=22cm,height=22cm,keepaspectratio=true]{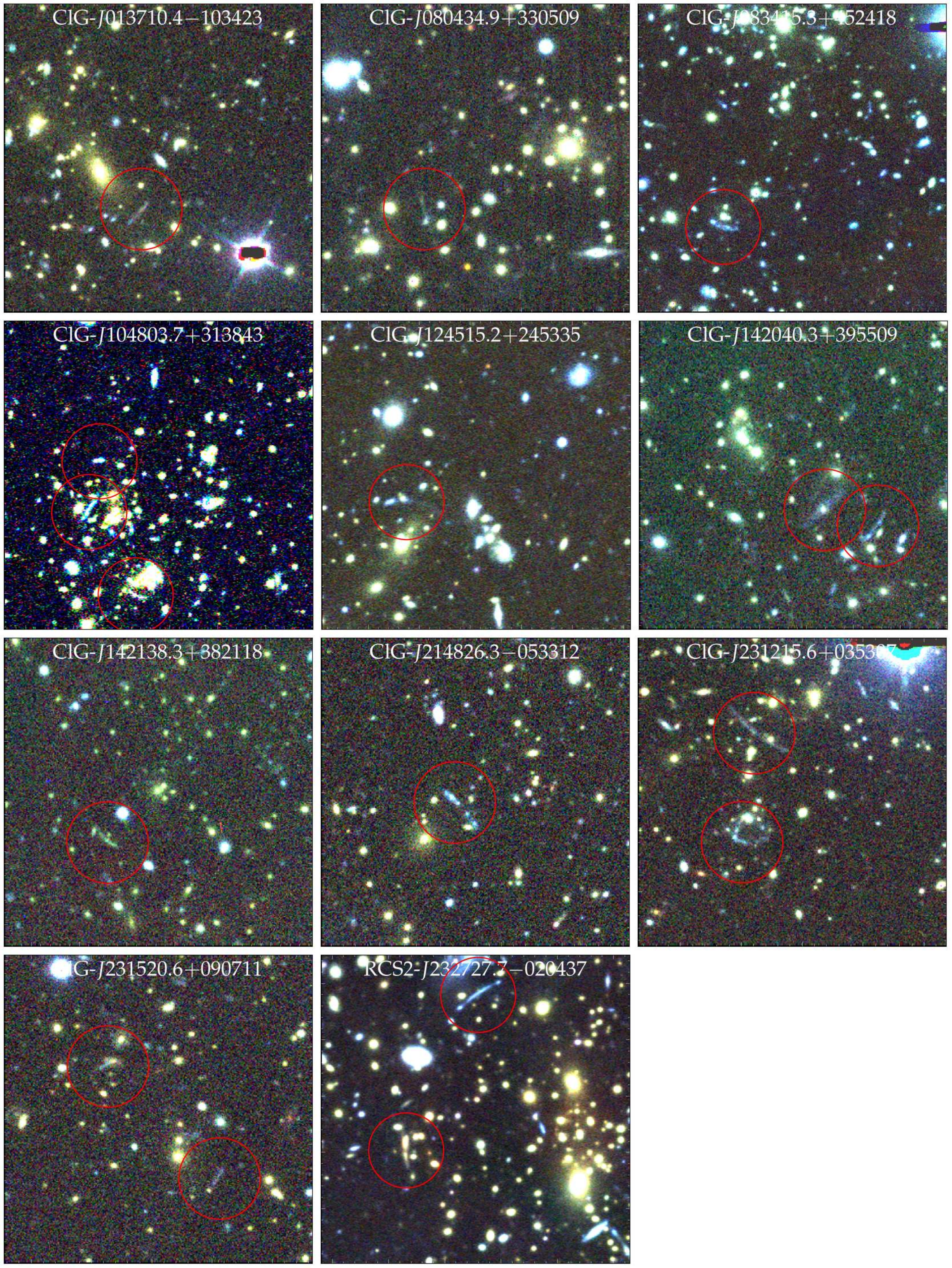}
 \caption{Strong lensing arc candidates. All panels show $75\arcsec\times75\arcsec$. Arc candidates are 
          highlighted by a red circle. }
 \label{fig:arcs}
\end{figure*}

By inspecting the colour images, 11 clusters with one or more potential strong gravitational lensing 
features were found. Those clusters and the arc coordinates are listed in Table \ref{tab:arcs} and corresponding 
colour images can be found in Fig. \ref{fig:arcs}. 

Due to the two clusters RCS2-$J$232727.7$-$020437 and ClG-$J$120958.9$+$495352 being in both the WHT as well as 
the LBT sample, we have the 
possibility to cross-check the results. The red sequence redshifts both agree within $2\sigma$ 
with the spectroscopic redshift. Comparing richness, we see that for ClG-$J$120958.9$+$495352 the measured 
values from the WHT 
sample are, within the error bars, consistent with the ones from the LBT sample (ClG-$J$120958.9$+$495352: 
$18\pm5;\; 22\pm5$). For RCS2-$J$232727.7$-$020437 the Schechter function fit did not work for the LBT data and 
thus the estimate for $N_{\rm{gal}}=11\pm6$ is very different to the one from the WHT ($46\pm7$). 
This is due to the values we fix the parameters in the Schechter function to. 
Those apparently do not match the observed data for RCS2-$J$232727.7$-$020437 in the deeper LBT data. 

\changes{Six of the clusters in this sample had been discovered independently by 
\citet{wen}\changesthree{, another} four by the \textit{Planck} collaboration \citep{planck_sz_cut}. 
We marked those clusters in Table \ref{tab:rich_red}. }

\begin{table}
\caption{This table names clusters where potential strong lensing features were found and gives their coordinates. }
\begin{tabular}{|l|c|c}
\hline \hline
Object & \multicolumn{1}{l|}{RA} & \multicolumn{1}{l|}{Dec} \\ \hline
ClG-$J$013710.4$-$103423 & 01:37:09.87 & $-$10:34:31.15 \\ \hline
ClG-$J$080434.9$+$330509 & 08:04:37.90 & $+$33:04:53.49 \\ \hline
ClG-$J$083415.3$+$452418 & 08:34:16.82 & $+$45:23:24.15 \\ \hline
ClG-$J$104803.7$+$313843 & 10:48:04.68 & $+$31:38:51.70 \\ 
 & 10:48:03.71 & $+$31:38.29.46 \\ 
 & 10:48:04.47 & $+$31:39:05.18 \\ \hline
ClG-$J$124515.2$+$245335 & 12:45:15.25 & $+$24:53:46.61 \\ \hline
ClG-$J$142040.3$+$395509 & 14:20:37.48 & $+$39:54:48.53 \\ 
 & 14:20:38.61 & $+$39:54:52.47 \\ \hline
ClG-$J$142138.3$+$382118 & 14:21:39.41 & $+$38:21:05.21 \\ \hline
ClG-$J$214826.3$-$053312 & 21:48:25.77 & $-$05:33.02.26 \\ \hline
ClG-$J$231215.6$+$035307 & 23:12:16.79 & $+$03:52:38.90 \\ 
 & 23:12:16.99 & $+$03:52:12.15 \\ \hline
ClG-$J$231520.6$+$090711 & 23:15:21.73 & $+$09:07:34.09 \\ 
 & 23:15:19.88 & $+$09:07:06.59 \\ \hline
RCS2-$J$232727.7$-$020437 & 23:27:29.41 & $-$02:03:48.03  \\ 
 & 23:27:30.69 & $-$02:04:29.47 \\ \hline
\end{tabular}
\label{tab:arcs}
\end{table}

\section{SZ Data Analysis}
\label{sec:sz}

The SZE signal is quantified in terms of the Compton $y$ parameter,
the line-of-sight integrated pressure. For scaling with mass, a
convenient measure is the integrated Comptonization
\begin{equation}
  Y = \int y \, \mathrm{d}\, \Omega = \frac{1}{(D_{A})^{2}} \frac{\sigma_{T}}{m_{e}c^2} \int \mathrm{d}l \int P(r) \mathrm{d}A,
\end{equation}
where $\Omega$ is the subtended solid angle of the cluster on the sky,
$D_{A}$ is the angular diameter distance, $\sigma_T$ is the Thomson
cross-section, $P(r)$ is the projected pressure profile and $A$ is a
projected physical area. Following \citet{marrone}, we quantify
the SZ signal in terms of the \emph{spherical} measure
\begin{equation}
  Y_{\mathrm{SZ}} \equiv Y_{\mathrm{sph}}(D_A)^{2} = \frac{\sigma_T}{m_ec^2} \int P(r) \mathrm{d}V,
\end{equation}
where $\mathrm{d}V$ is a physical volume element and $P(r)$ is now the
pressure as a function of physical radius. Note that we have moved
$D_A$ to the left-hand side of the equation to remove the redshift
dependence in the SZE measure.

For the pressure as a function of radial distance, we adopt the
generalized NFW pressure profile \citep{nagai}, with the
functional form
\begin{equation}
  P(r) = \frac{P_0}{c_{500} x^{\gamma}(1+c_{500}x^{\alpha})^{(\beta-\gamma)/\alpha}},
\end{equation}
where $x = r/r_{500}$ and ($P_0,c_{500},\alpha,\beta,\gamma$) are
parameters of the model. For our analysis, we fix
($\alpha,\beta,\gamma$) to the best-fitting values of the `universal
pressure profile' found by \citet{arnaud}.

We reduce the CARMA data using a pipeline similar to the one used in \citet{mucho}, which was adapted 
for the use with CARMA. We first filter out bad weather errors as well as pointing errors and then apply a 
gain and flux calibration. For the flux calibration we use the model of Mars from \citet{rudy87}. We assume that
Mars is a disc of uniform brightness, Fourier transform this disc to the visibility plane and compare it to the 
measured visibilities. From this comparison, we derive an antenna-specific scale factor, which brings the observations in
line with the model. A conservative estimate for the absolute flux calibration uncertainty is $\sim$7 per cent. This 
results from $\sim$5 per cent uncertainty in the model from \citet{rudy87} and $\sim$5 per cent uncertainty 
from the gain solution of the telescopes. 

We carry out a model fit using the pressure profile of
\citet{arnaud} to the interferometric data by Fourier transforming
the model and comparing it to the data in visibility space. We
minimize a $\chi^{2}$ statistic and estimate the detection significance.
If this significance is greater than three we estimate the spherical
volume-integrated Comptonization, $Y_{\rm{SZ}}$. If the significance
is less than three we only give upper limits on $Y_{\rm{SZ}}$ and the
mass. We call these cases non-detections. 
We estimate $r_{500}$ by forcing $Y_{\rm{SZ}}$ to be consistent with
the $Y_{SZ}-M_{500}$ scaling relation of \citet{andersson}, which effectively means we
are fitting only to integrated Comptonization (or equivalently, mass)
from which $r_{500}$ is directly given. We use the scaling relation with a 
fixed slope of 1.79. 
The positions and peak fluxes of point sources detected in the
long-baseline image are included in the fit (rather than subtracted in
the visibility-plane), and marginalized over in determining $Y_{\rm{SZ}}$.

In addition to the statistical errors in the fit there are further sources of
uncertainty. First, there is intrinsic scatter in the
$M-Y_\mathrm{SZ}$ scaling relation, for which we assume a 21 per cent
intrinsic scatter in mass consistent with \citet{andersson}. We
add this scatter in quadrature to the statistical errors of the fit as
it assumes that the clusters follow the scaling relation exactly.  In
addition, it is important to realize that this scaling relation has
been calibrated via the $M-Y_\mathrm{X}$ scaling relation, which
itself was calibrated empirically using weak lensing data at much lower
redshifts only \citep{viki}.  Given the high-redshift range of our clusters,
any deviation from the assumed self-similar redshift evolution would
lead to a systematic bias in the derived masses.  So far, \citet{jee} 
present the only weak lensing study for a large cluster
sample at high redshifts. Their analysis suggests a possible evolution
in the $M-T_\mathrm{X}$ scaling relation until \mbox{$z\sim 1$} in
comparison to self-similar evolution at the \mbox{$20-30\%$} level.
To be conservative, and accounting for the in comparison to \citet{jee}
slightly lower redshift range of our clusters
(\mbox{$z_\mathrm{median}=0.725$}), we therefore adopt an additional 20 per cent
systematic uncertainty in the mass scale.
\citet{andersson} use a cosmology slightly
different to ours, introducing another systematic bias of about 5
per cent in mass, which is however negligible compared to the statistical errors. 

For ClG-$J$122208.6$+$422924, which was observed in a different configuration, we used the 6-m and 10-m antennas to 
search for point sources and the 3.5-m antennas to estimate $Y_{SZ}$. 
We analysed about 4 hours of these data but could not 
detect the cluster. Half of the data had only been observed at half the normal bandwidth. 

\begin{figure}
\centering
 \includegraphics[width=8.5cm,height=8.5cm,keepaspectratio=true]{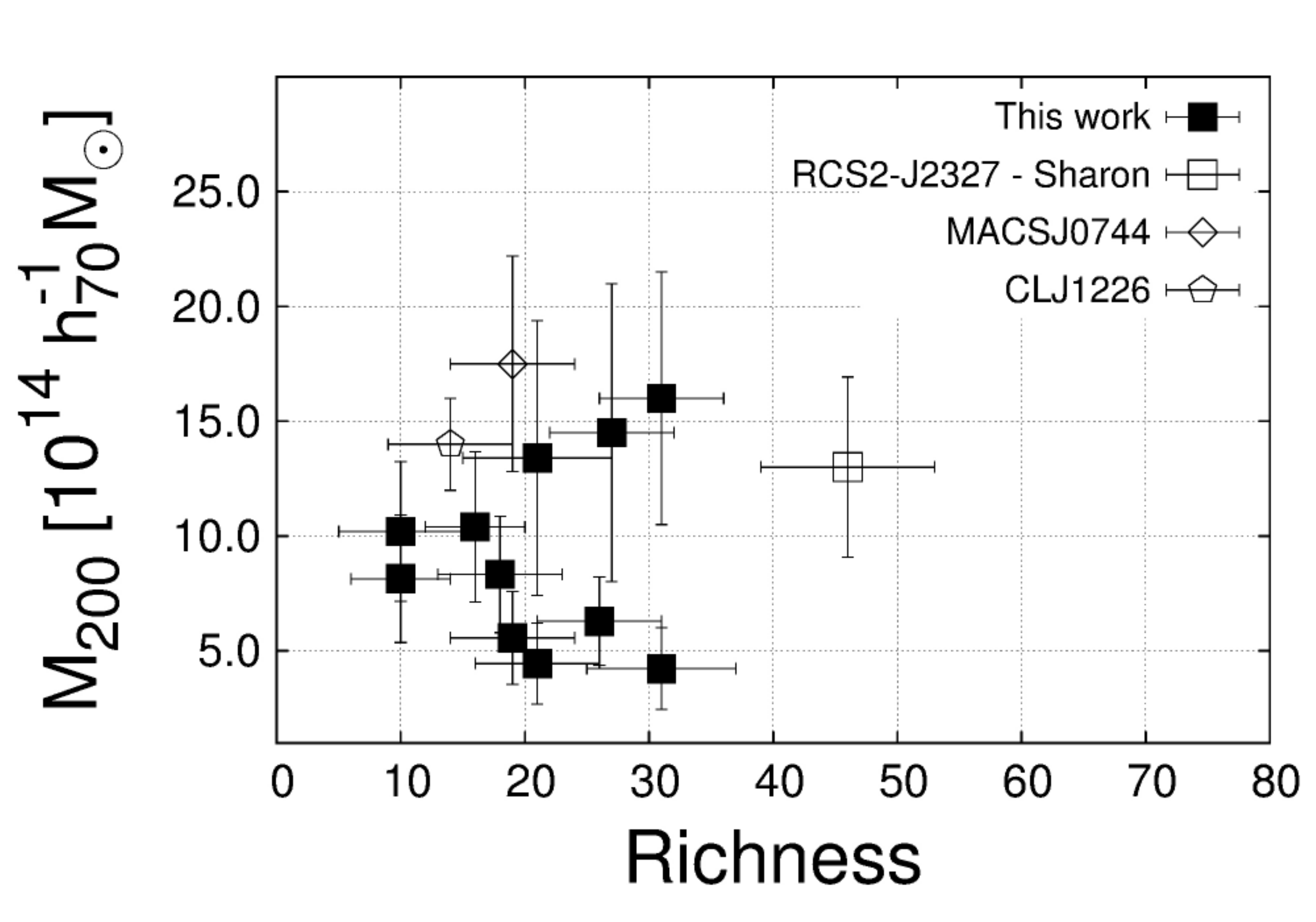}
 \caption{We show mass estimates as a function of richness. The solid points are SZ masses from this study.  
          The open symbols are masses from previous studies. 
          The masses for \mbox{RCS2-$J$232727.7$-$020437} were determined from $Y_{SZ}$ given in Sharon et al. (in prep.), 
          which was measured from 
          CARMA data. For MACS074452.8$+$392725 the mass estimate is taken from \citet{umetsu}, which is 
          a weak lensing mass estimate. \citet{clj_mass} measure a weak lensing 
          mass for ClG$J$1226+33. 
          The error bars in mass for objects from this work include the 21 per cent scatter from 
          the scaling relation from \citet{andersson}
          but not the 20 per cent systematic error due to the high-redshift mass calibration (see Section 
          \ref{sec:sz}).}
 \label{fig:mass_richness}
\end{figure}

From the 21 clusters analysed we detect 11. For those we estimate
$M_{500}$ according to the scaling relation. Furthermore, using the
mass-concentration relation from \citet{duffy} we can convert this to
$M_{200}$. Again, for the non-detections, we only determine upper
limits. In \mbox{Fig. \ref{fig:mass_richness}} we show how the masses
from the SZ data scale with our richness estimates. Additionally, we
also show masses which were already known for
RCS2-$J$232727.7$-$020437, MACS074452.8$+$392725
and ClG$J$1226+33. 
$M_{200}$ for RCS2-$J$232727.7$-$020437 was determined from the value given for $Y_{SZ}$ in Sharon et al. 
(in prep.), which had been measured from CARMA data. 
\changesthree{We estimate 
\mbox{$M_{200}=(11.3\pm3.9)\times 10^{14}\,h^{-1}_{70}\,\rm{M_{\odot}}$} using the cosmology adopted in our work; 
the given uncertainty is dominated by the uncertainties in the scaling relation. }
For MACS074452.8$+$392725 we use the weak lensing mass estimate from \citet{umetsu}. 
Also, \citet{clj_mass} estimate a weak lensing mass for ClG$J$1226+33. 
The mass estimates for MACS074452.8$+$392725 and ClG$J$1226+33 use different techniques than we do, 
which means that they do not necessarily measure the same mass as our SZ estimate. 

In the plot there is only a rough relation between mass and richness visible;
one can see large scatter among the data. This is expected due to
comparably short integration times, the assumptions we make while
determining the masses but most importantly due to the large intrinsic
scatter between mass and richness (e.g. \citealt{angulo}).
We also find that our $M_{500}$ estimates range mostly between
$3-9\times10^{14}h^{-1}_{70}\rm{M_{\odot}}$ at redshifts of $0.6\leq
z\leq0.9$. That we only find these high masses is due to a selection
effect; the less massive clusters could not be detected at $>3\sigma$
in the SZ data while using only these comparably short integration
times.

The objects that have not been detected with CARMA are in most cases
not particularly rich in the optical or were only integrated for a
short amount of time. There are two exceptions.  One of these is
ClG-$J$142040.3$+$395509, for which we find a point source at the BCG
position, which can potentially cancel the SZ-signal. Due to a flagged
antenna, we do not have enough long baselines to properly measure the
flux of this source. This could explain the apparent strong SZ-peak,
with an offset of about $2\arcmin$ from the BCG position.  The other one is
\mbox{ClG-$J$095416.5$+$173808}, which is optically rich, but not
detected. As we already explained before, there is a large scatter in
the mass-richness relation, so this could mean that
\mbox{ClG-$J$095416.5$+$173808} shows a strong richness while not being
massive, which would result in a faint SZ signal.

\begin{figure}
\centering
 \includegraphics[width=8.5cm,height=8.5cm,keepaspectratio=true]{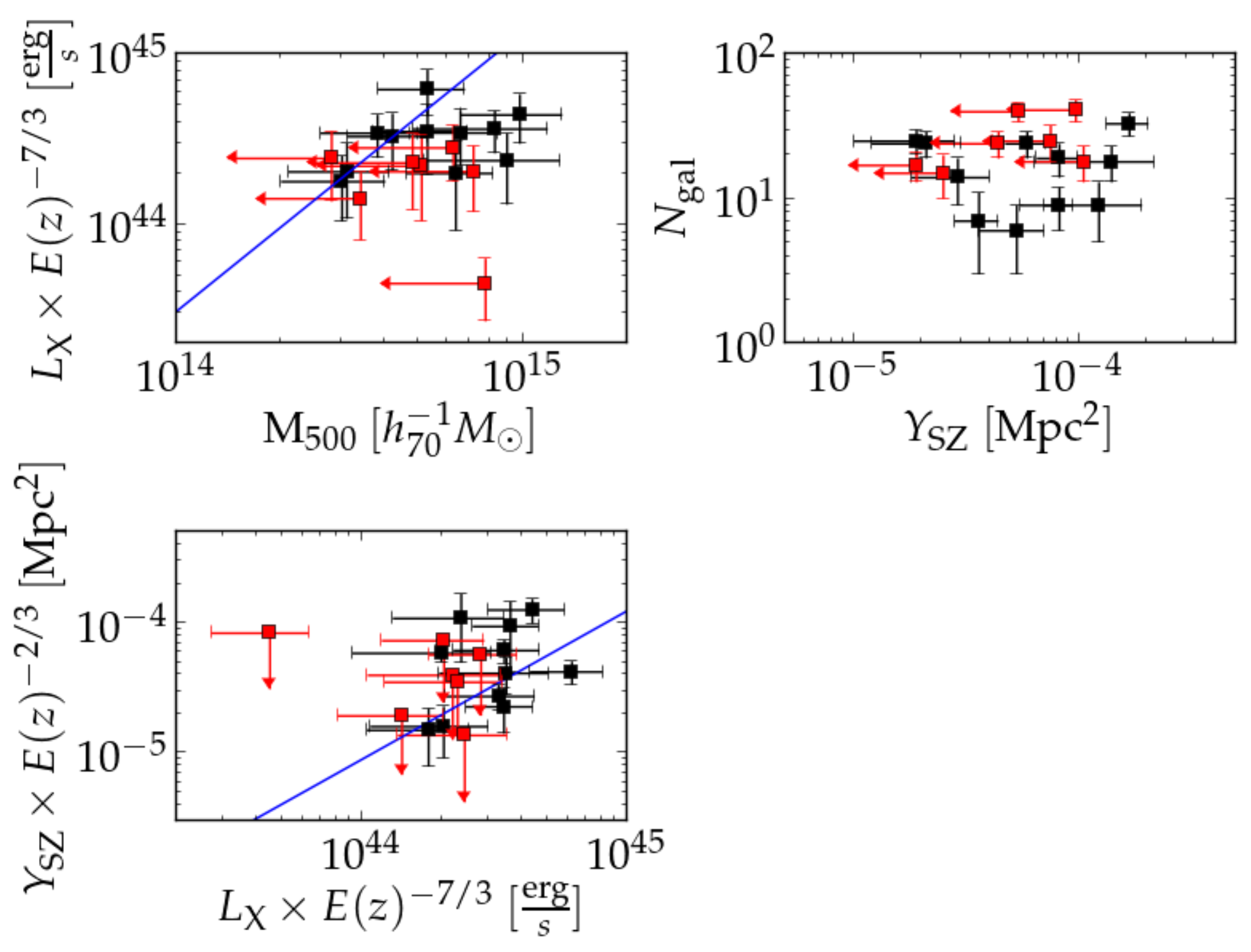}
 \caption{\changes{We present scaling relations comparing the \textit{ROSAT} X-ray luminosity $L_{\rm{X}}$, the 
          integrated Comptonization parameter $Y_{\rm{SZ}}$ from the CARMA data, 
          the SZ-inferred galaxy cluster mass $M_{500}$, and the 
          cluster richness $N_{\rm{gal}}$. The black points show \changestwo{the CARMA detections}, 
          the red points the CARMA non-detections, and the corresponding $3\sigma$ upper limits. 
          The blue lines show corresponding relations from \citet{arnaud}. 
          \changestwo{We assume self-similar evolution in order to compare the data to
          the scaling relations from \citet{arnaud}. }
          For a detailed discussion please see Section \ref{sec:sz}\changesthree{.}
          }
          }
 \label{fig:scalings}
\end{figure}

\changes{In Fig. \ref{fig:scalings}, we show the \changestwo{$M_{500}-L_{\mathrm{X}}$}, the 
$L_{\mathrm{X}}-Y_{\rm{SZ}}$, and the $Y_{\rm{SZ}}-N_{\rm{gal}}$ scaling relations. The blue lines show 
the corresponding \changestwo{$M_{500}-L_{\mathrm{X}}$} and $L_{\mathrm{X}}-Y_{\rm{SZ}}$ relations from \citet{arnaud}. 
\changestwo{In order to compare the data to those relations, we assume self-similar evolution, which depends 
on the self-similar evolution factor 
\mbox{$E(z)=H(z)/H_{0}=\sqrt{\Omega_{\mathrm{m}}(1+z)^{3}+\Omega_{\Lambda}}$} (in this form it is only true for flat
cosmologies). 
}
We plot both, the CARMA detections, as well as the non-detections 
(denoted in red) using their $3\sigma$ upper limits. The measured \changestwo{$M_{500}-L_{\mathrm{X}}$} and the 
$L_{\mathrm{X}}-Y_{\rm{SZ}}$ relations agree well with the results from \citet{arnaud}. 
\changestwo{The non-detections seem to have a preferentially lower $L_{\mathrm{X}}$ than the detection}. 
When comparing $N_{\rm{gal}}$ to $Y_{\rm{SZ}}$ we find no clear trend, as already discussed for 
\mbox{Fig. \ref{fig:mass_richness}}. 
\changestwo{We do not attempt to compare the mass-richness or Ysz-richness relations to previous works, due to differences
in the definition of richness between studies. }
}

All results from the CARMA SZ observations can be found in Table
\ref{tab:carma}.  In addition to the CARMA data we also check if the
clusters observed with CARMA can be found in data from
\textit{Planck}. A detailed description of this and postage stamps of
the CARMA and \textit{Planck} SZ-maps are given in the appendix.

\section{Are there galaxy clusters too massive compared to predictions from $\Lambda$CDM?}
\label{sec:outliers}

Using $M_{200}$ estimated from the SZ data, we can check for 11 clusters if they are too massive for 
our current structure formation paradigm. For this we use the fitting formula given in \citet{mortonson} for upper 
mass limits as a function of redshift and survey size in a flat $\Lambda$CDM cosmology. One limitation here is 
that we do not test the whole sample but every cluster individually. 
We do not know the exact area which has been used for our cluster detection, due to our selection procedure. 
Nevertheless, we can calculate a lower limit for the area. For this we use all galaxies from the SDSS DR8, which 
includes the complete SDSS imaging data, with 
$\mathtt{psfMag\_i}<13$ and all objects from the RASS faint source catalogue. 
We grid both samples and compute the overlapping area as the sum of cells, which contain at least one object 
of each survey. This estimate does strongly depend on the cell size and does not converge. 
In order to find a lower limit on the area used, we vary the cell size and check how many of the 44 clusters are within
the overlapping area. The smallest cell size for which we still find all clusters within the overlap is 
$0.7\times0.7\,\rm{deg^{2}}$. For this configuration, we find the area to be $\approx10,000\,\rm{deg^{2}}$. 
This estimate is, as mentioned before, only a lower limit and it does not take variations in sensitivity in the SDSS
and RASS into account. Thus, we only provide this area estimate to put our findings into a cosmological context. We also 
test if our sample selection is sensitive to the exposure time in RASS. We find the lowest exposure time of a 
cluster in the sample to be $\approx350\,\rm{s}$. Areas in RASS with exposure times greater or equal to 
these $350\,\rm{s}$ correspond to about 80 per cent of the total RASS area. 

\begin{figure}
\centering
 \includegraphics[width=8.5cm,height=8.5cm,keepaspectratio=true]{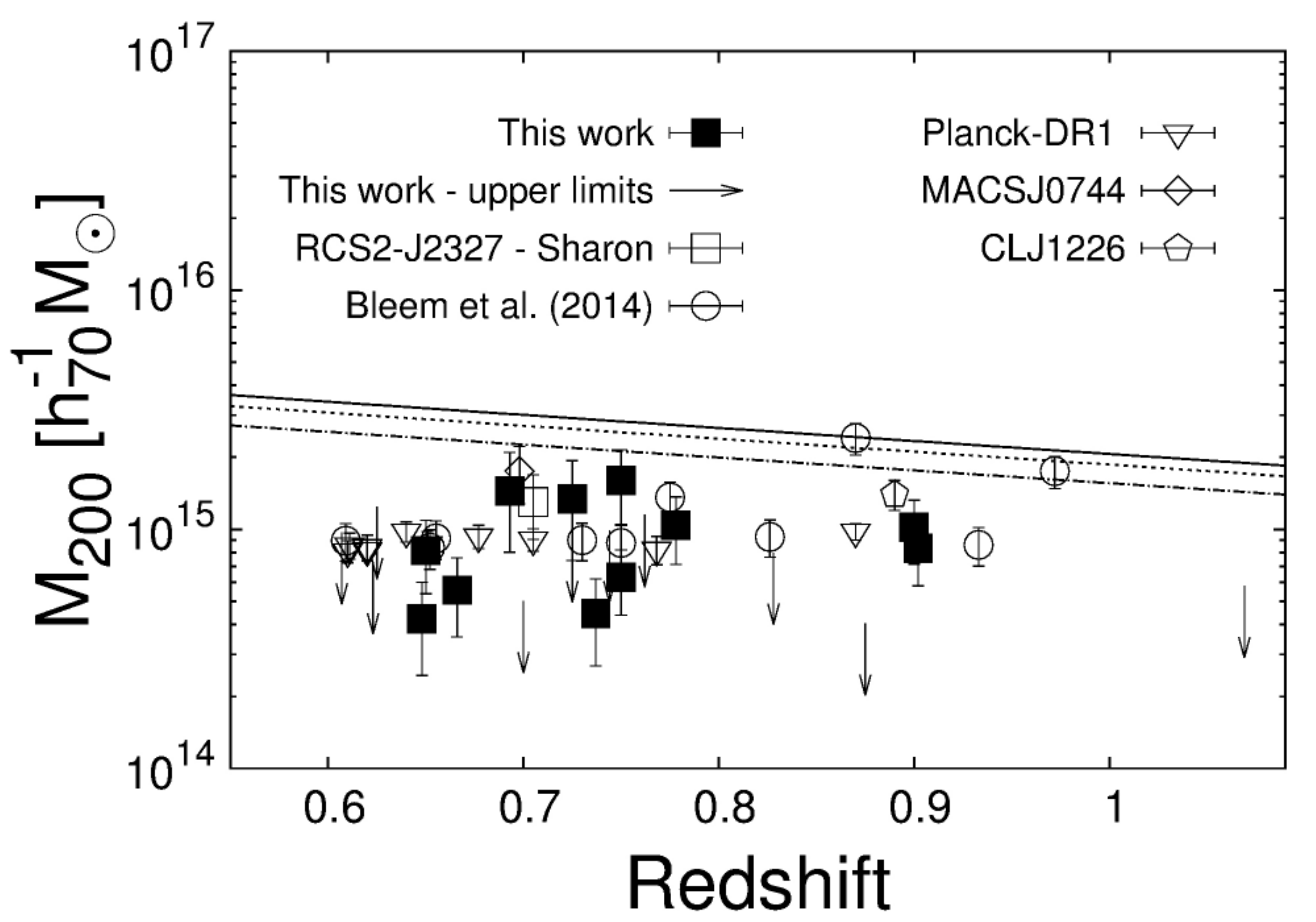}
 \caption{The solid line shows the 99 per cent confidence mass limit as a function of redshift 
          for a flat $\Lambda$CDM universe and 
          the survey size of \textit{Planck}. The dotted line shows the same limit for a survey size of 
          $10,000\,\rm{deg^{2}}$, which corresponds to the survey size in this work. The dash-dotted one 
          shows the corresponding limit for the SPT $2500\,\rm{deg^{2}}$ survey. To compute these 
          lines we use the fitting formula from \citet{mortonson}  and acknowledge the fact that this 
          gives too strict limits. The solid points show the masses estimated in this study. The other
          symbols represent masses from previous studies. Arrows indicate the upper limits we find in CARMA for 
          non-detected clusters. We find no tension with the $\Lambda$CDM model. The open 
          circles are the 10 most massive clusters between $0.6<z<1.0$ from \citet{bleem}, the triangles 
          with the tip down show the 10 most massive clusters in this redshift range
          in the \textit{Planck} SZ sample \citep{planck_sz}. 
          }
 \label{fig:mass_limits}
\end{figure}
We plot the cluster masses against redshift in \mbox{Fig. \ref{fig:mass_limits}}. 
Additionally, the masses of three clusters from previous studies
are plotted (see Section \ref{sec:sz}). 
Furthermore, we take the 10 most massive clusters at redshifts $0.6<z<1.0$ from 
\citet{bleem}\footnote{\url{http://pole.uchicago.edu/public/data/sptsz-clusters/index.html}} 
and also from \textit{Planck} 
\citep{planck_sz}\footnote{\url{http://pla.esac.esa.int/pla/aio/planckProducts.html}}, determine their 
$M_{200}$ as described above and plot them as well. The masses we find for both of these samples 
are comparable to ours. Considering that we use the most massive ones from that study, this might again be an
indication of the massive and extreme nature of our cluster sample. 

As visible in Fig. \ref{fig:mass_limits} we find no significant tension between our clusters and the current cosmological 
standard model. The clusters from \citet{bleem} and from \textit{Planck} were found by different surveys using a 
different selection 
function. Thus, from Fig. \ref{fig:mass_limits} we should not infer possible tension for those clusters. 
We are aware that \changes{\citet{hotchkiss}} showed that the fitting formula we use is too strict, but since 
none of the objects is in strong tension, the method from \citet{mortonson} is sufficient for our purposes. 

Only a sub-sample is tested here and ideally we would like to achieve mass estimates for more clusters than these 11, 
preferably for those with the highest $N_{\rm{gal}}$, since this should be a rough indication for the mass.

\section{Notes on Individual Clusters}
\label{sec:outstanding}
After this summary of the general data, we will now focus on the most notable objects in the sample, which are
either high-redshift clusters, very rich clusters in the optical or very massive clusters according to their SZ signal. 
Those clusters are the most interesting targets
for further and deeper follow-up observations, in order to determine their masses and other interesting properties.

\subsection{RCS2-$J$232727.7$-$020437}
We can confirm this object to be a very rich cluster. The measured photometric 
redshift \mbox{$z_{\rm{phot}}=0.725\pm0.042$} agrees well with the known spectroscopic one,  
\mbox{$z_{\rm{spec}}=0.705$} \citep{2327}. 
The richness 
of \mbox{$46\pm7$} (WHT) is the largest in the sample, as expected from its high mass. However, as we will see 
later on, 
there are comparable clusters in our sample. Its mass has been estimated before. For example 
\citet{gralla} find \mbox{$M_{500}=(6.2\pm0.8)\times 10^{14}\,h^{-1}_{70}\,\rm{M_{\odot}}$}.  
\citet{hasselfield} estimate masses from ACT data and use different ways to fix their $Y-M$ scaling relation. 
This leads to different mass estimates for RCS2-$J$232727.7$-$020437. The results from using what they call a 
'universal pressure profile' (UPP) scaling relation is \mbox{$M_{500}=(9.4\pm1.5)\times10^{14}\,h^{-1}_{70}\rm{M_{\odot}}$}, which
is their lowest estimate. 
Their largest value arises from using a scaling relation determined with dynamical masses. With 
\mbox{$M_{500}=(14.9\pm3.0)\times10^{14}\,h^{-1}_{70}\rm{M_{\odot}}$} 
this is about 50 per cent larger than the UPP value. 
From the $Y_{SZ}$ given in Sharon et al. (in prep.)
we estimate \mbox{$M_{500}=(8.1\pm2.3)\times 10^{14}\,h^{-1}_{70}\,\rm{M_{\odot}}$}. This agrees well with the value from 
\citet{gralla} and within $1\sigma$ with the UPP mass from \citet{hasselfield}.

\subsection{ClG-$J$095416.5$+$173808}
ClG-$J$095416.5$+$173808 has a measured photometric redshift of \mbox{$z_{\rm{phot}}=0.725\pm0.047$}, 
which scatters $2\sigma$ low compared to \mbox{$z_{\rm{spec}}=0.828$} \citep{0954}. 
The richness of \mbox{$N_{\rm{gal}}=40\pm6$} 
is comparable to the one of RCS2-$J$232727.7$-$020437. 
We do not detect this object with more than $3\sigma$ in CARMA. 

\subsection{ClG-$J$104803.7$+$313843}
This object is located at \mbox{$z_{\rm{phot}}=0.750\pm0.047$} and has a richness of \mbox{$31\pm5$}. Its redshift 
appears to be slightly higher than 
the one measured for RCS2-$J$232727.7$-$020437. 
The existence of at least two potential arcs indicates a high mass, which is confirmed from the SZ observations, 
where we estimate \mbox{$M_{200}=(16.0\pm5.5)\times10^{14}\,h^{-1}_{70}\rm{M_{\odot}}$}. This makes it one of the most massive 
systems known at high redshift. 

\subsection{ClG-$J$120958.9$+$495352}
We discovered ClG-$J$120958.9$+$495352, which has a spectroscopic
redshift of \mbox{$z_{\rm{spec}}=0.902$} and a measured red sequence redshift of 
\mbox{$z_{\rm{phot}}=0.950\pm0.112$} (WHT). 
The richness 
within 0.5 Mpc was measured to be \mbox{$18\pm5$} (WHT). Due to its large distance we cannot probe the
luminosity function down to faint magnitudes. Based on the RASS count rate this is the most X-ray luminous cluster
discovered by our programme with $L_{\rm{X}}=(20.3\pm6.2)\times10^{44}\frac{\rm{erg}}{\rm{s}}$. 
In addition to the findings from the optical data, we estimate its SZ-mass to be
\mbox{$M_{200}=(8.3\pm2.5)\times10^{14}\,h^{-1}_{70}\rm{M_{\odot}}$}. 

\subsection{ClG-$J$122208.6$+$422924}
\label{sec:1222}
ClG-$J$122208.6$+$422924 is the object with the highest measured red sequence redshift in our sample 
(\mbox{$z_{\rm{phot}}=1.000\pm0.200$}). 
The large error 
arises from the fact that only very few cluster members are visible, and those have an average $i$-band
magnitude of \mbox{$i\approx24.1$}, which is very close to the detection limit. The measured spectroscopic 
redshift is somewhat larger with \mbox{$z_{\rm{spec}}=1.069$}, which was measured from the two brightest cluster
members, and both spectra show a clear break at the corresponding $4000\,\rm{\AA}$ position. 
This makes the object by far the highest redshift one in the sample. Nevertheless, the two brightest 
cluster galaxies are detected in the SDSS, which given their high-redshift is very rare and might 
indicate a high mass for those galaxies. 
We do not detect this object in $4\,\rm{h}$ of CARMA data but measure a $3\sigma$ upper mass 
limit $M_{500}<3.8\times 10^{14}\,h^{-1}_{70}\rm{M_{\odot}}$.

\subsection{ClG-$J$133732.5$+$195827}
This cluster has a redshift of $z_{\rm{phot}}=0.900\pm0.106$, but it was observed at a high airmass, which 
might have affected the data. It does show a strong SZ signal, and considering its possibly high redshift 
its mass of \mbox{$M_{200}=(10.2\pm3.0)\times10^{14}h^{-1}_{70}\rm{M_{\odot}}$} 
is extraordinarily high. The optical colour image shows only a few very red galaxies, and we measure its richness 
as $N_{\rm{gal}}=10\pm5$. 

\subsection{ClG-$J$135345.0$+$432905}
ClG-$J$135345.0$+$432905 shows a very strong SZ signal and with $M_{200}=(13.4\pm6.0)\times10^{14}h^{-1}_{70}\rm{M_{\odot}}$ 
it is among the most massive clusters in the CARMA sample. It has no spectroscopic redshift but we measure the
red sequence redshift to $z_{\rm{phot}}=0.725\pm0.024$. Its richness is $N_{\rm{gal}}=21\pm6$. 

\subsection{ClG-$J$142040.3$+$395509}
This cluster has a spectroscopic redshift of \mbox{$z_{\rm{spec}}=0.607$} \citep{red} and shows a richness of 
\mbox{$N_{\rm{gal}}=25\pm5$}. 
From serendipitous \textit{Chandra} observations, we conducted an X-ray analysis, which can be found in 
Appendix \ref{app:x}. 
This analysis shows a gas temperature of about $8^{+3}_{-2}$ keV, which indicates a high mass. Also, \citet{red} find 
several strong lensing features and a
high velocity dispersion of \mbox{$\sigma_{v}=1095^{+86}_{-175}\rm{\frac{km}{s}}$}. 
\changes{\citet{oguri} use weak and strong gravitational lensing to measure its virial mass, 
which, adapted to the cosmology we use, is \mbox{$M_{\mathrm{vir}}=10.77^{+3.59}_{-2.88}\times10^{14}\rm{M_{\odot}}$}. }
Still, we did not detect 
this cluster at more than $3\sigma$ using CARMA. A possible explanation is a point source we find at the
BCG position. This source could counter act the SZ signal and thus we would not detect 
the cluster. 
\mbox{ClG-$J$142040.3$+$395509} shows a strong signal in \textit{Planck}. 

\subsection{ClG-$J$142138.3$+$382118}
With a measured redshift of \mbox{$z_{\rm{phot}}=0.750\pm0.027$} (\mbox{$z_{\rm{spec}}=0.762$}) and a richness of 
\mbox{$41\pm7$}, this cluster appears to be at higher redshift but with a comparable richness 
to RCS2-$J$232727.7$-$020437.  
Possible strong lensing arcs have been observed which also indicate a high mass. 
On the other hand we cannot detect it at more than $3\sigma$ in the CARMA data, which could be due to the short 
integration time of only 1.3 hours. 

\subsection{ClG-$J$152741.9$+$204443}
ClG-$J$152741.9$+$204443 has a redshift of \mbox{$z_{\rm{spec}}=0.693$} and a richness of 
$N_{\rm{gal}}=27\pm5$. This is a rather large richness, which also agrees with the CARMA analysis. There we find one
of the strongest SZ signals, which corresponds to a mass of $M_{200}=(14.5\pm6.5)\times10^{14}h^{-1}_{70}\rm{M_{\odot}}$. 
Again, this appears to be an exceptionally massive cluster.

\section{Conclusions}
\label{sec:conclusio}

We cross-correlated RASS and SDSS in order to find rich galaxy clusters at redshifts 
\mbox{$0.6\lesssim z \lesssim 1.0$}. 
Using follow-up observations we confirmed 44 cluster candidates. 
The motivation was to find similar objects as \mbox{RCS2-$J$232727.7$-$020437}, in which we succeeded. 
We estimated red sequence redshifts which we compared to our spectroscopic sub-sample and
determined the cluster richness by fitting and integrating a Schechter function. 

In the end, we found at least two clusters of comparable richness as RCS2-$J$232727.7$-$020437. 
Furthermore, we achieved rough mass estimates from SZ observations for a sub-sample of 11 clusters and find them 
to be massive systems. Using the formalism by \citet{mortonson} we find no tension between any of these clusters 
and the standard cosmological model.  
Further investigations, which will need deeper and higher-quality observations, will reveal the masses of more of 
these rare objects and check whether those are compatible with the $\Lambda$CDM structure formation paradigm. 

We have demonstrated that the approach of cross-correlating X-ray with optical data within an area of about 
10,000 $\rm{deg^{2}}$ is efficient
resulting in the discovery of some of the richest galaxy clusters at high-$z$ to date. 

Our cluster sample is unique and complementary to \changes{the \textit{Planck} cluster sample and also
to the Southern hemisphere samples of the SPT and ACT. With respect to the redshift range and 
the large area this sample is more similar to the \textit{Planck} sample than to the other two. }
Although we have constructed our sample by surveying a large area, we cannot attempt to infer cosmological parameters 
from it. The sample is by construction incomplete, because we searched for the most massive objects, 
which are easiest to detect.

\section*{Acknowledgements}

A.B. was supported for this research partly through a stipend from the International Max Planck
Research School (IMPRS) for Astronomy and Astrophysics at the Universities of Bonn and
Cologne and through funding from the Transregional Collaborative Research Centre (TR 33) of the DFG. \\
\changestwo{CHG is supported by NSF grant AST-1140019.} \\
T.S. acknowledges support from the German Federal Ministry of Economics and Technology (BMWi) provided 
through DLR under project 50 OR 1308. \\
H.H. is supported by the DFG Emmy Noether grant Hi 1495/2-1. \\
R.G.M. is supported in part by the U.S. Department of Energy under contract 
number DE-AC02-76SF00515. \\
T.H.R. acknowledges support from the DFG through the Heisenberg research grant RE 1462/5, 
from grant RE 1462/6, and from TR 33 (project B18). \\
The Dark Cosmology Centre is funded by the Danish National Research Foundation. 

We would like to thank Douglas Applegate and \mbox{Matthias} Klein for fruitful discussions about the optical data and 
Chris Benn for his help with the ACAM data. 

We thank Dominik Klaes and Inka Hammer for conducting early test observations for our project with the 1-m telescope 
of Bonn University at the Hoher List observatory as well as Patrick Kelly, who conducted test observations with the 
2.1-m KPNO telescope. Also, we would like to thank Bas Nefs, who conducted test observations with the Isaac Newton Telescope. 

We also thank Tom Plagge for his initial analysis of the CARMA data. 

\changes{
While preparing this paper we made use of the python version of Ned Wright's cosmology calculator 
\citep{cosmo_calc}, which was implemented by James Schombert. 
Additionally, we used \texttt{TOPCAT} \citep{topcat} and \texttt{STILTS} \citep{stilts}. 

For preparing this work we made use of the following facilities and archives: the William Herschel Telescope, 
the Large Binocular Telescope, the Combined Array for Research in Millimeter Astronomy, the SDSS Archive, 
the \textit{ROSAT} bright and faint source catalogs, the \textit{Chandra} archive, and the 
\textit{Planck} legacy archive. }

\small
Author contributions:
AB led the optical reduction, redshift and richness analysis, cosmological comparison, and paper writing.
TS conceived and led the overall programme and the optical follow-up campaign.
RGM led the {\it \textit{Chandra}} analysis.
The CARMA observations, reductions, and analysis were conducted by CHG, MSo and DM.
JE created the $y$-maps from the \textit{Planck} data. 
HHo, WH, AM, MSc, TE, HHi, AvdL contributed to the optical observations, reductions, and data analysis.
THR and AH contributed to the further multi-wavelength follow-up campaign.
All authors have contributed to the writing of this paper.

\normalsize

\bibliography{mylib}

\appendix

\section{X-ray Analysis of ClG-$J$142040.3$+$395509}
\label{app:x}
\begin{figure*}
 \centering
 \includegraphics[width=18cm,height=15cm,keepaspectratio=true]{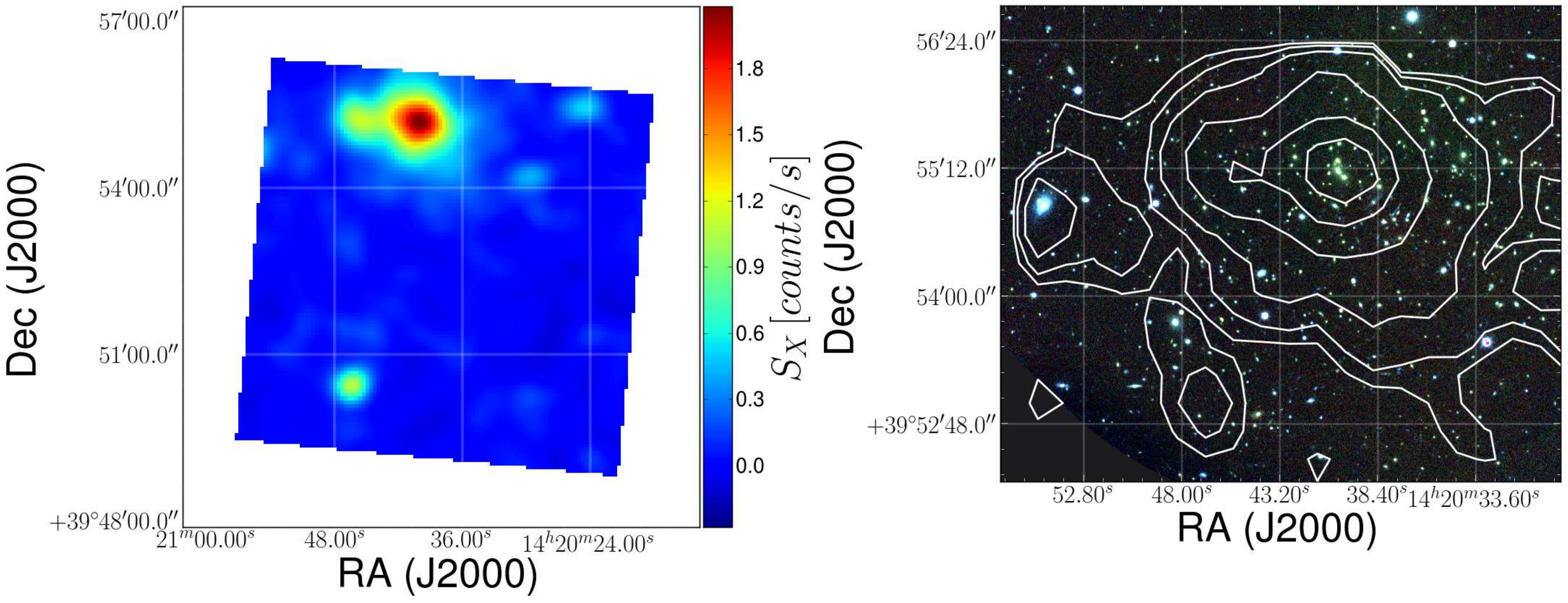}
 \caption{left: smoothed \textit{Chandra} image of ClG-$J$142040.3$+$395509. Colour indicates the count rate. 
          The cluster is visible in the upper part of the image. right: 
          white lines are X-ray contours from \textit{Chandra} superposed on the optical three-colour
          image from the WHT. This image 
          shows a much smaller part of the field than visible in the left-hand panel, due to the
          small field of view in the optical. The contour levels are 0.05, 0.1, 0.2, 0.5, 1.0, and 1.5 counts/s. }
 \label{fig:xray}
\end{figure*}

In addition to our optical and SZ data, we found serendipitous archival data from \textit{Chandra} for
\mbox{ClG-$J$142040.3$+$395509}.
Fig. \ref{fig:xray} shows the \mbox{$0.6-7.0\,\rm{keV}$} count rate image. This was background subtracted and 
exposure corrected. When fitting a simple \mbox{free absorption + thermal Bremsstrahlung} model to the data
we find the metallicity to be $0.5^{+0.4}_{-0.3}\,Z_{\odot}$ and a temperature of $T_{X}=8^{+3}_{-2}$ keV. This 
temperature as well as the high velocity dispersion from \citet{red} indicate a high mass. 
The flux in the 0.6-7.0 keV band is $S_{X}=6.8\times 10^{-13} \rm{\frac{erg}{cm^{2}\,s}}$. 
The 2-10 keV luminosity is $L_{X}=8.1\times10^{44}\rm{\frac{erg}{s}}$ which is consistent with the 
RASS luminosity in Table \ref{tab:rich_red}. In Fig. \ref{fig:xray} we also show the optical three-colour
image superposed with the X-ray contours. Clearly, the X-ray peak coincides with the position of the BCG. 

The chip, which is being analysed here is acis-S4, which is non-standard for the analysis of extended sources. 
This means that the calibration model is probably not as reliable as normal which might result in an
additional systematic bias of our measurements. 
All errors given are $1\sigma$ errors.

\section{Galaxy Cluster and SZ Data}

\begin{landscape}
\begin{table}\scriptsize
\caption{This table shows photometric and spectroscopic redshifts, richness, 
        X-ray counts in $\rm{1/s}$ 
        and the X-ray luminosity in $10^{44}\rm{\frac{erg}{s}}$ measured in
        the \textit{ROSAT} $0.1-2.4\rm{keV}$ band. $N_{\rm{gal}}$ is the richness measured using the Schechter 
        function fit, $N_{\rm{count}}$ the one measured by counting galaxies brighter than $m^{*}+2$. 
        In the comments column we have noted several anomalies, namely: $^1$ bimodal red sequence 
        galaxy 
        distribution as a function of redshift or only a weak signal (compare Fig. 
        \ref{fig:2312_all} top-right panel); 
        $^2$ the data are shallow compared to other images; $^3$ the object had been observed 
        at high 
        airmass ($\approx 2$); $^4$ the Schechter function fit did not work well, which means 
        that our fixed 
        parameters lead to poor fits. For clusters with known spectroscopic 
        redshift we used those 
        instead of the 
        photometric ones for the richness estimate. Values in brackets indicate the LBT values 
        for clusters which 
        have data from both telescopes. Coordinates given are those of the BCG. 
        \changes{Cluster names showing the superscript $^{\dagger}$ were independently discovered by 
        \citet{wen}; the superscript $^{\diamond}$ indicates clusters that have been independently 
        discovered by the \textit{Planck} collaboration \changestwo{\citep{planck_sz_cut}}. } }
\begin{tabular}{|l|c|c|c|c|c|c|c|c|c|c|c|c|}
\hline \hline
Object & WHT & LBT & SZ & Ra & Dec & $z_{\rm{phot}}$ & $z_{\rm{spec}}$ & $N_{\rm{gal}}$ & $N_{\rm{count}}$ & $\rm{counts}_{X}$ & $L_{X}$ & com.\\ \hline
ClG-$J$001640.6$-$130644\changes{$^{\dagger}$} & x & - & - & 00:16:40.636 & $-$13:06:43.84 & 0.700$\pm$0.097 & - & 18$\pm$6 & 13$\pm$4 & 0.0297$\pm$0.0126 & 6.5$\pm$3.1 & $^{1} $ \\ \hline
ClG-$J$005805.6$+$003058 & x & - & - & 00:58:05.648 & $+$00:30:57.85 & 0.725$\pm$0.017 & - & 33$\pm$6 & 39$\pm$6 & 0.0230$\pm$0.0099 & 5.5$\pm$2.4 & $^{} $ \\ \hline
ClG-$J$013710.4$-$103423 & x & - & - & 01:37:10.433 & $-$10:34:23.15 & 0.525$\pm$0.034 & 0.662 & 24$\pm$5 & 17$\pm$4 & 0.0192$\pm$0.0091 & 2.9$\pm$1.4 & $^{4} $ \\ \hline
ClG-$J$031924.2$+$404055 & x & - & - & 03:19:24.237 & $+$40:40:54.91 & 0.750$\pm$0.020 & 0.680 & 6$\pm$3 & 13$\pm$4 & 0.0229$\pm$0.0087 & 5.0$\pm$1.9 & $^{1,4} $ \\ \hline
MACS$J$074452.8$+$392725 & x & - & - & 07:44:52.775 & $+$39:27:25.45 & 0.675$\pm$0.028 & 0.698 & 19$\pm$5 & 17$\pm$4 & 0.0148$\pm$0.0073 & 3.1$\pm$1.5 & $^{} $ \\ \hline
ClG-$J$080434.9$+$330509\changes{$^{\dagger \diamond}$} & x & - & - & 08:04:34.899 & $+$33:05:08.99 & 0.575$\pm$0.014 & 0.552 & 24$\pm$5 & 26$\pm$5 & 0.0239$\pm$0.0102 & 3.0$\pm$1.3 & $^{} $ \\ \hline
ClG-$J$083415.3$+$452418\changes{$^{\dagger}$} & - & x & x & 08:34:15.317 & $+$45:24:18.19 & 0.675$\pm$0.029 & 0.666 & 19$\pm$5 & 13$\pm$4 & 0.0416$\pm$0.0116 & 8.1$\pm$2.3 & $^{} $ \\ \hline
ClG-$J$084009.8$+$442154 & x & - & - & 08:40:09.783 & $+$44:21:53.51 & 0.700$\pm$0.060 & - & 21$\pm$4 & 26$\pm$5 & 0.0689$\pm$0.0209 & 15.0$\pm$5.0 & $^{2} $ \\ \hline
ClG-$J$093503.2$+$061438 & x & - & x & 09:35:03.235 & $+$06:14:38.46 & 0.750$\pm$0.031 & - & 9$\pm$3 & 8$\pm$3 & 0.0435$\pm$0.0135 & 11.2$\pm$3.6 & $^{} $ \\ \hline
ClG-$J$094742.3$+$351742 & - & x & - & 09:47:42.313 & $+$35:17:41.81 & 0.500$\pm$0.065 & - & 20$\pm$4 & 2$\pm$1 & 0.0190$\pm$0.0086 & 1.8$\pm$0.9 & $^{4} $ \\ \hline
ClG-$J$094700.0$+$631905 & x & - & - & 09:47:00.010 & $+$63:19:04.99 & 0.700$\pm$0.062 & 0.710 & 9$\pm$4 & 10$\pm$3 & 0.0172$\pm$0.0076 & 3.8$\pm$1.8 & $^{4} $ \\ \hline
ClG-$J$094811.6$+$290709 & x & - & x & 09:48:11.569 & $+$29:07:09.48 & 0.775$\pm$0.063 & 0.778 & 16$\pm$4 & 21$\pm$5 & 0.0337$\pm$0.0109 & 9.5$\pm$3.4 & $^{} $ \\ \hline
ClG-$J$095416.5$+$173808 & - & x & - & 09:54:16.461 & $+$17:38:07.76 & 0.725$\pm$0.047 & 0.828 & 40$\pm$6 & 15$\pm$4 & 0.0216$\pm$0.0110 & 6.5$\pm$3.4 & $^{} $ \\ \hline
ClG-$J$102714.5$+$034500 & x & - & - & 10:27:14.475 & $+$03:45:00.36 & 0.700$\pm$0.030 & 0.749 & 20$\pm$5 & 33$\pm$6 & 0.0304$\pm$0.0126 & 7.4$\pm$3.1 & $^{} $ \\ \hline
ClG-$J$103605.6$+$441140 & x & - & - & 10:36:05.645 & $+$44:11:40.29 & 0.800$\pm$0.119 & - & 5$\pm$3 & 13$\pm$4 & 0.0228$\pm$0.0100 & 6.9$\pm$3.6 & $^{4} $ \\ \hline
ClG-$J$104803.7$+$313843 & - & x & x & 10:48:03.669 & $+$31:38:42.90 & 0.750$\pm$0.047 & - & 31$\pm$5 & 10$\pm$3 & 0.0443$\pm$0.0132 & 12.1$\pm$3.9 & $^{} $ \\ \hline
ClG-$J$120958.9$+$495352 & x & x & x & 12:09:58.948 & $+$49:53:52.02 & 0.950(0.925)$\pm$0.112(0.027) & 0.902 & 18(22)$\pm$5(5) & 16(40)$\pm$4(6) & 0.0486$\pm$0.0145 & 20.3$\pm$6.2 & $^{} $ \\ \hline
ClG-$J$122208.6$+$422924 & x & - & x & 12:22:08.612 & $+$42:29:24.19 & 1.000$\pm$0.200 & 1.069 & 7$\pm$4 & 12$\pm$3 & 0.0189$\pm$0.0087 & 11.3$\pm$7.2 & $^{4} $ \\ \hline
Cl$J$1226$+$33 & x & - & - & 12:26:58.170 & $+$33:32:48.41 & 0.950$\pm$0.028 & 0.892 & 14$\pm$5 & 27$\pm$5 & 0.0233$\pm$0.0102 & 9.5$\pm$4.2 & $^{} $ \\ \hline
ClG-$J$124515.2$+$245335\changes{$^{\dagger}$} & - & x & x & 12:45:15.204 & $+$24:53:35.41 & 0.650$\pm$0.027 & - & 10$\pm$4 & 15$\pm$1 & 0.0233$\pm$0.0100 & 4.2$\pm$1.8 & $^{} $ \\ \hline
ClG-$J$131104.8$+$551443 & x & - & - & 13:11:04.815 & $+$55:14:42.67 & 0.775$\pm$0.127 & - & 12$\pm$5 & 10$\pm$3 & 0.0204$\pm$0.0080 & 5.7$\pm$2.8 & $^{} $ \\ \hline
ClG-$J$131339.7$+$221151 & x & - & - & 13:13:39.723 & $+$22:11:50.82 & 0.675$\pm$0.019 & 0.737 & 21$\pm$5 & 25$\pm$5 & 0.0232$\pm$0.0106 & 5.3$\pm$2.5 & $^{} $ \\ \hline
ClG-$J$133620.3$+$544540 & - & x & x & 13:36:20.308 & $+$54:45:40.22 & 0.875$\pm$0.039 & - & 17$\pm$4 & 20$\pm$4 & 0.0204$\pm$0.0088 & 7.7$\pm$3.4 & $^{} $ \\ \hline
ClG-$J$133732.5$+$195827 & x & - & x & 13:37:32.451 & $+$19:58:26.57 & 0.900$\pm$0.106 & - & 10$\pm$5 & 11$\pm$3 & 0.0162$\pm$0.0079 & 6.5$\pm$3.5 & $^{3} $ \\ \hline
ClG-$J$135345.0$+$432905\changes{$^{\diamond}$} & x & - & x & 13:53:44.996 & $+$43:29:05.12 & 0.725$\pm$0.024 & - & 21$\pm$6 & 42$\pm$6 & 0.0393$\pm$0.0107 & 9.3$\pm$2.6 & $^{} $ \\ \hline
\end{tabular}
\label{tab:rich_red}
\end{table}
\end{landscape}

\begin{landscape}
\begin{table}\scriptsize
\contcaption{}
\begin{tabular}{|l|c|c|c|c|c|c|c|c|c|c|c|c|}
\hline \hline
Object & WHT & LBT & SZ & Ra & Dec & $z_{\rm{phot}}$ & $z_{\rm{spec}}$ & $N_{\rm{gal}}$ & $N_{\rm{count}}$ & $\rm{counts}_{X}$ & $L_{X}$ & com.\\ \hline
ClG-$J$142008.8$-$031906 & x & - & - & 14:20:08.763 & $-$03:19:06.40 & 0.750$\pm$0.070 & - & 16$\pm$4 & 20$\pm$4 & 0.0209$\pm$0.0112 & 5.4$\pm$3.0 & $^{} $ \\ \hline
ClG-$J$142040.3$+$395509\changes{$^{\dagger \diamond}$} & x & - & x & 14:20:40.353 & $+$39:55:09.72 & 0.600$\pm$0.044 & 0.607 & 25$\pm$5 & 34$\pm$6 & 0.0360$\pm$0.0100 & 5.5$\pm$1.6 & $^{} $ \\ \hline
ClG-$J$142138.3$+$382118\changes{$^{\diamond}$} & x & - & x & 14:21:38.288 & $+$38:21:18.32 & 0.750$\pm$0.027 & 0.762 & 41$\pm$7 & 42$\pm$6 & 0.0209$\pm$0.0084 & 5.5$\pm$2.3 & $^{2} $ \\ \hline
ClG-$J$142227.4$+$233739 & x & - & - & 14:22:27.366 & $+$23:37:38.82 & 0.750$\pm$0.017 & 0.726 & 23$\pm$5 & 20$\pm$4 & 0.0284$\pm$0.0105 & 6.9$\pm$2.6 & $^{} $ \\ \hline
ClG-$J$143411.9$+$175039 & x & - & x & 14:34:11.929 & $+$17:50:38.96 & 0.800$\pm$0.020 & 0.744 & 25$\pm$7 & 35$\pm$6 & 0.0278$\pm$0.0101 & 7.4$\pm$2.7 & $^{3} $ \\ \hline
ClG-$J$144847.4$+$284312 & x & - & - & 14:48:47.381 & $+$28:43:12.17 & 0.750$\pm$0.125 & - & 3$\pm$2 & 3$\pm$2 & 0.0186$\pm$0.0074 & 4.8$\pm$2.4 & \changestwo{$^{1,2,3}$} \\ \hline
ClG-$J$145508.4$+$320028 & x & - & - & 14:55:08.384 & $+$32:00:27.90 & 0.675$\pm$0.017 & 0.654 & 11$\pm$4 & 27$\pm$5 & 0.0161$\pm$0.0068 & 3.0$\pm$1.3 & $^{} $ \\ \hline
ClG-$J$150532.2$+$331249\changes{$^{\dagger}$} & x & - & - & 15:05:32.212 & $+$33:12:48.83 & 0.725$\pm$0.036 & 0.757 & 24$\pm$5 & 24$\pm$5 & 0.0142$\pm$0.0058 & 3.6$\pm$1.5 & $^{} $ \\ \hline
ClG-$J$151544.3$+$042554 & x & - & - & 15:15:44.312 & $+$04:25:53.65 & 0.700$\pm$0.039 & - & 32$\pm$8 & 29$\pm$5 & 0.0262$\pm$0.0100 & 5.7$\pm$2.2 & $^{} $ \\ \hline
ClG-$J$151601.9$+$394426 & x & - & x & 15:16:01.946 & $+$39:44:26.57 & 0.725$\pm$0.025 & - & 10$\pm$5 & 34$\pm$6 & 0.0288$\pm$0.0084 & 6.8$\pm$2.0 & $^{2,3,4} $ \\ \hline
ClG-$J$152741.9$+$204443 & x & - & x & 15:27:41.933 & $+$20:44:42.77 & 0.700$\pm$0.034 & 0.693 & 27$\pm$5 & 14$\pm$4 & 0.0269$\pm$0.0122 & 5.8$\pm$2.6 & $^{} $ \\ \hline
ClG-$J$153035.0$+$130512 & x & - & - & 15:30:34.980 & $+$13:05:12.31 & 0.625$\pm$0.025 & - & 17$\pm$4 & 34$\pm$6 & 0.0216$\pm$0.0103 & 3.6$\pm$1.7 & $^{3,4} $ \\ \hline
ClG-$J$153258.8$+$021324 & x & - & - & 15:32:58.807 & $+$02:13:23.87 & 0.850$\pm$0.119 & - & 13$\pm$3 & 12$\pm$3 & 0.0209$\pm$0.0086 & 7.3$\pm$3.6 & $^{1,3} $ \\ \hline
ClG-$J$153735.6$+$382851 & x & - & x & 15:37:35.582 & $+$38:28:50.90 & 0.750$\pm$0.057 & - & 26$\pm$5 & 29$\pm$5 & 0.0336$\pm$0.0116 & 8.7$\pm$3.2 & $^{} $ \\ \hline
ClG-$J$171225.8$+$561253 & x & - & - & 17:12:25.840 & $+$56:12:52.51 & 0.600$\pm$0.036 & - & 19$\pm$5 & 13$\pm$4 & 0.0082$\pm$0.0036 & 1.2$\pm$0.6 & $^{} $ \\ \hline
ClG-$J$174109.9$+$555819 & x & - & x & 17:41:09.881 & $+$55:58:19.06 & 0.625$\pm$0.036 & - & 18$\pm$5 & 27$\pm$5 & 0.0059$\pm$0.0023 & 1.0$\pm$0.4 & $^{} $ \\ \hline
ClG-$J$214826.3$-$053312 & x & - & - & 21:48:26.270 & $-$05:33:12.01 & 0.625$\pm$0.025 & - & 23$\pm$5 & 34$\pm$6 & 0.0211$\pm$0.0095 & 3.5$\pm$1.6 & $^{} $ \\ \hline
ClG-$J$223007.6$-$080949 & x & - & x & 22:30:07.589 & $-$08:09:48.80 & 0.575$\pm$0.017 & 0.623 & 24$\pm$5 & 25$\pm$5 & 0.0336$\pm$0.0154 & 5.1$\pm$2.4 & $^{} $ \\ \hline
ClG-$J$223727.5$+$135523 & x & - & x & 22:37:27.543 & $+$13:55:22.56 & 0.700$\pm$0.043 & - & 15$\pm$5 & 24$\pm$5 & 0.0161$\pm$0.0067 & 3.5$\pm$1.5 & $^{} $ \\ \hline
ClG-$J$231215.6$+$035307 & x & - & x & 23:12:15.600 & $+$03:53:06.90 & 0.625$\pm$0.030 & 0.648 & 31$\pm$6 & 19$\pm$4 & 0.0234$\pm$0.0093 & 4.1$\pm$1.7 & $^{} $ \\ \hline
ClG-$J$231520.6$+$090711 & x & - & - & 23:15:20.558 & $+$09:07:11.09 & 0.725$\pm$0.024 & - & 20$\pm$5 & 36$\pm$6 & 0.0206$\pm$0.0086 & 4.9$\pm$2.1 & $^{} $ \\ \hline
RCS2-$J$232727.7$-$020437 & x & x & - & 23:27:27.7 & $-$02:04:37.00 & 0.725(0.650)$\pm$0.042(0.024) & 0.705 & 46(11)$\pm$7(6) & 35(35)$\pm$6(6) & 0.0474$\pm$0.0137 & 10.7$\pm$3.6 & $^{4\rm{(LBT)}} $ \\ \hline
\end{tabular}
\label{tab:rich_red}
\end{table}
\end{landscape}

\begin{table*}
\caption{This table shows the results from the SZ observations. $t_{\rm{int}}$ is the integration 
               time in hours. All masses are
               in $10^{14}h^{-1}_{70}\rm{M_{\odot}}$, $Y_{\rm{SZ}}$ is given in $10^{-5}\rm{Mpc^{2}}$. 
               To the statistical mass errors from the fit, we have added in quadrature the 21 per cent 
               scatter from the scaling relation from \citet{andersson}. We acknowledge the fact that
               our error bars do not include the 20 per cent systematic error from the uncertainty 
               in the high-redshift mass calibration (see Section \ref{sec:sz}).}
\begin{tabular}{l|c|c|c|c|c|c}
\hline \hline
Object & $z_{\rm{spec}}$ & $z_{\rm{phot}}$ & $Y_{SZ}$ & $M_{500}$ & $M_{200}$ & $t_{\rm{int}}$ \\ \hline
ClG-$J$083415.3$+$452418 & 0.666 & 0.675 & 2.9$\pm$1.1 & 3.8$\pm$1.2 & 5.6$\pm$2.0 & 3.5 \\ \hline
ClG-$J$094811.6$+$290709 & 0.778 & 0.775 & 8.2$\pm$1.7 & 6.6$\pm$1.9 & 10.4$\pm$3.3 & 4.1 \\ \hline
ClG-$J$095416.5$+$173808 & 0.828 & 0.725 & $<5.4$ & $<5.1$ & $<8.0$ & 4.0 \\ \hline
ClG-$J$104803.7$+$313843 & - & 0.750 & 16.8$\pm$3.6 & 9.8$\pm$3.2 & 16.0$\pm$5.5 & 3.5 \\ \hline
ClG-$J$120958.9$+$495352 & 0.902 & 0.950 & 5.9$\pm$1.2 & 5.3$\pm$1.5 & 8.3$\pm$2.5 & 6.8 \\ \hline
ClG-$J$122208.6$+$422924 & 1.069 & 1.000 & $<3.7$ & $<3.8$ & $<5.8$ & 4.0 \\ \hline
\changes{ClG-$J$124515.2$+$245335} & - & 0.650 & 5.3$\pm$1.7 & 5.3$\pm$1.6 & 8.1$\pm$2.8 & 4.1 \\ \hline
ClG-$J$131339.7$+$221151 & 0.737 & 0.675 & 2.1$\pm$0.9 & 3.1$\pm$1.0 & 4.4$\pm$1.8 & 5.8 \\ \hline
ClG-$J$133620.3$+$544540 & - & 0.875 & $<1.9$ & $<2.8$ & $<4.0$ & 3.6 \\ \hline
ClG-$J$133732.5$+$195827 & - & 0.900 & 8.2$\pm$1.2 & 6.4$\pm$1.8 & 10.2$\pm$3.0 & 1.4 \\ \hline
ClG-$J$135345.0$+$432905 & - & 0.725 & 12.3$\pm$6.8 & 8.3$\pm$3.5 & 13.4$\pm$6.0 & 1.6 \\ \hline
\changes{ClG-$J$142040.3$+$395509} & 0.607 & 0.600 & $<7.0$ & $<6.2$ & $<9.7$ & 1.5 \\ \hline
ClG-$J$142138.3$+$382118 & 0.762 & 0.750 & $<9.7$ & $<7.2$ & $<11.5$ & 1.3 \\ \hline
ClG-$J$143411.9$+$175039 & 0.744 & 0.800 & $<7.5$ & $<6.3$ & $<9.9$ & 1.7 \\ \hline
\changes{ClG-$J$151601.9$+$394426} & - & 0.725 & $<7.5$ & $<6.3$ & $<9.9$ & 0.9 \\ \hline
ClG-$J$152741.9$+$204443 & 0.693 & 0.700 & 14.0$\pm$7.6 & 9.0$\pm$3.8 & 14.5$\pm$6.5 & 1.8 \\ \hline
ClG-$J$153735.6$+$382851 & - & 0.750 & 3.6$\pm$0.8 & 4.2$\pm$1.1 & 6.3$\pm$1.9 & 5.1 \\ \hline
ClG-$J$174109.9$+$555819 & - & 0.625 & $<10.5$ & $<7.8$ & $<12.4$ & 4.5 \\ \hline
ClG-$J$223007.6$-$080949 & 0.623 & 0.575 & $<4.4$ & $<4.8$ & $<7.3$ & 3.4 \\ \hline
ClG-$J$223727.5$+$135523 & - & 0.700 & $<2.5$ & $<3.4$ & $<5.0$ & 6.3 \\ \hline
ClG-$J$231215.6$+$035307 & 0.648 & 0.625 & 1.9$\pm$0.9 & 3.0$\pm$1.0 & 4.2$\pm$1.8 & 6.8 \\ \hline
\end{tabular}
\label{tab:carma}
\end{table*}

\section{Postage Stamps of all Clusters}
\begin{figure*}
 \centering
 \includegraphics[width=16.875cm, height=22.5cm,keepaspectratio=true]{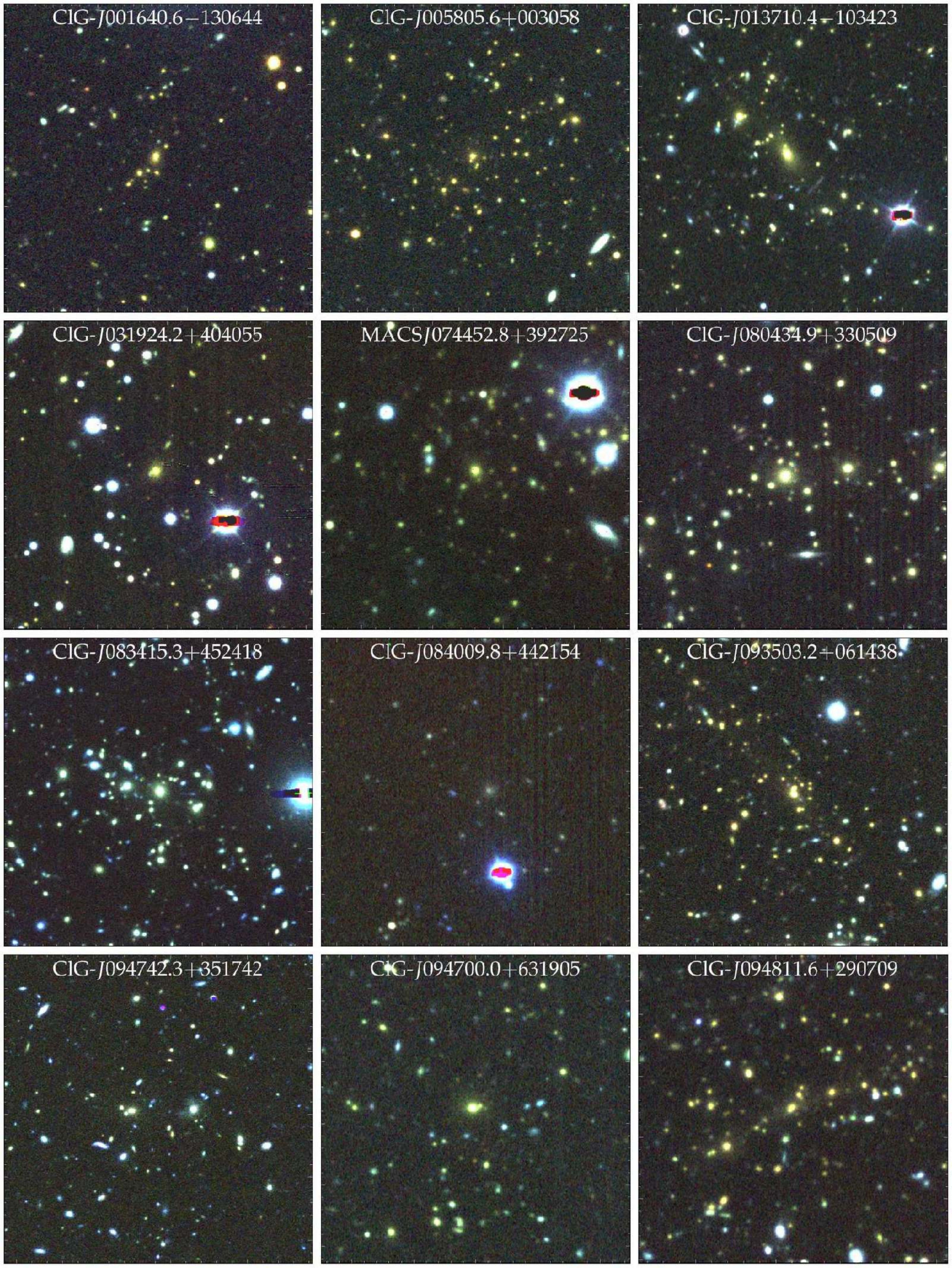}
 \caption{In this figure, we present optical postage stamps of all clusters in our sample. 
These postage stamps were created using the $r$-, $i$- and $z$ band images from WHT and LBT. Wherever 
available we show the LBT data, which is considerably deeper. Which data are available can be found in 
Table \ref{tab:rich_red}. All images show the inner $1\farcm7$ 
of the cluster. }
 \label{fig:opt_all1}
\end{figure*}

\begin{figure*}
 \centering
 \includegraphics[width=16.875cm, height=22.5cm,keepaspectratio=true]{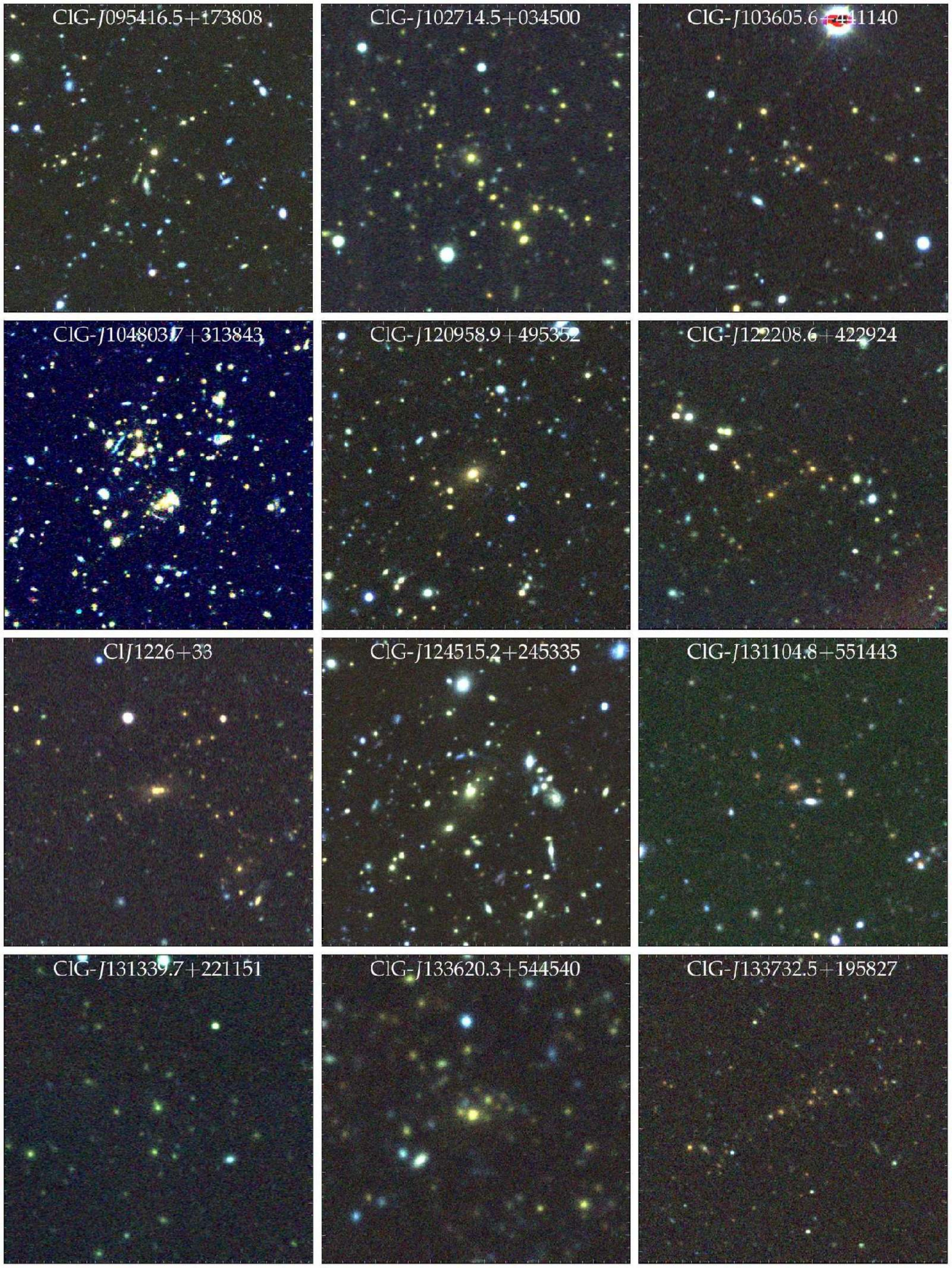}
 \contcaption{}
 \label{fig:opt_all2}
\end{figure*}

\begin{figure*}
 \centering
 \includegraphics[width=16.875cm, height=22.5cm,keepaspectratio=true]{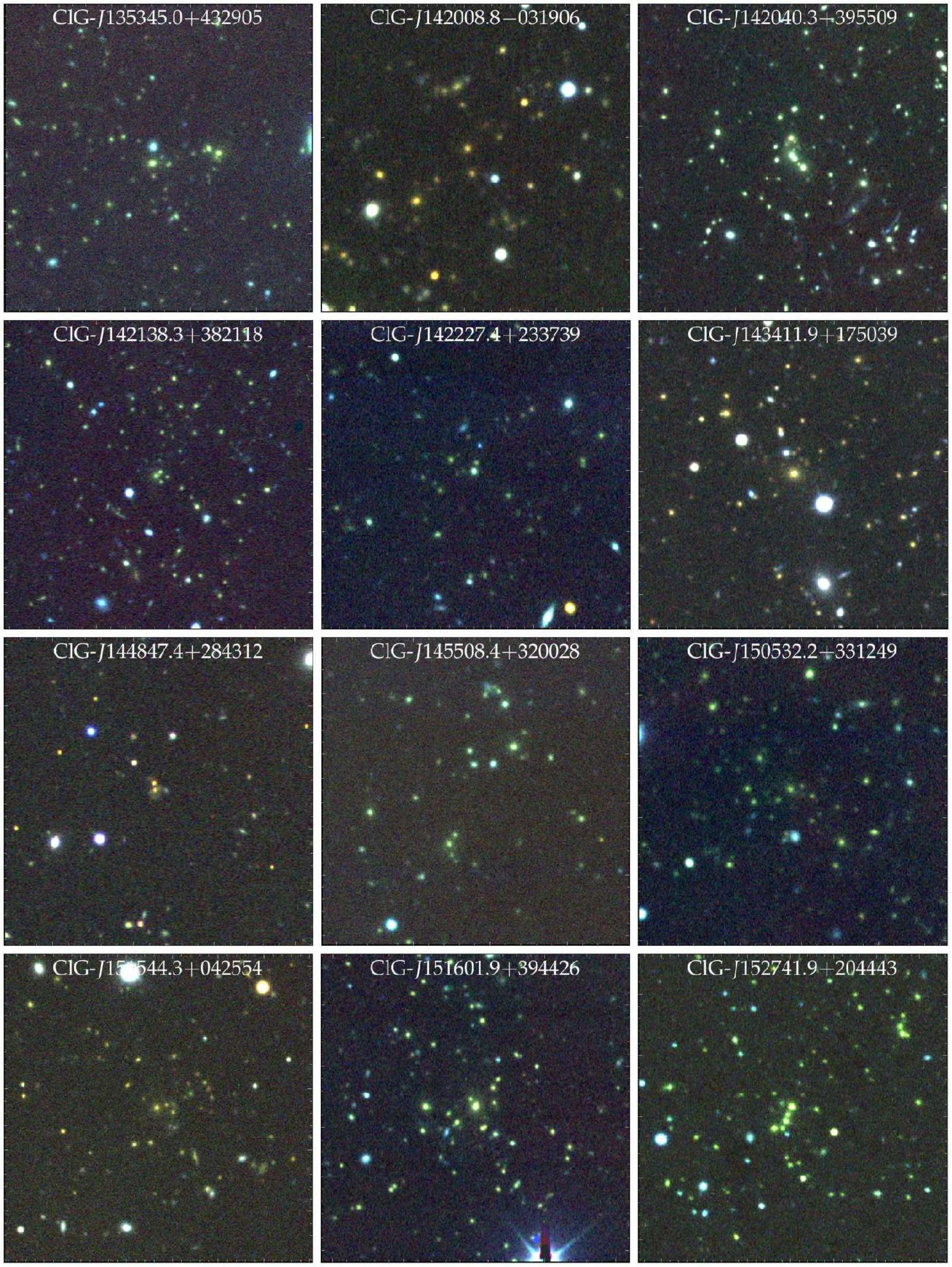}
 \contcaption{}
 \label{fig:opt_all3}
\end{figure*}

\begin{figure*}
 \centering
 \includegraphics[width=16.875cm, height=22.5cm,keepaspectratio=true]{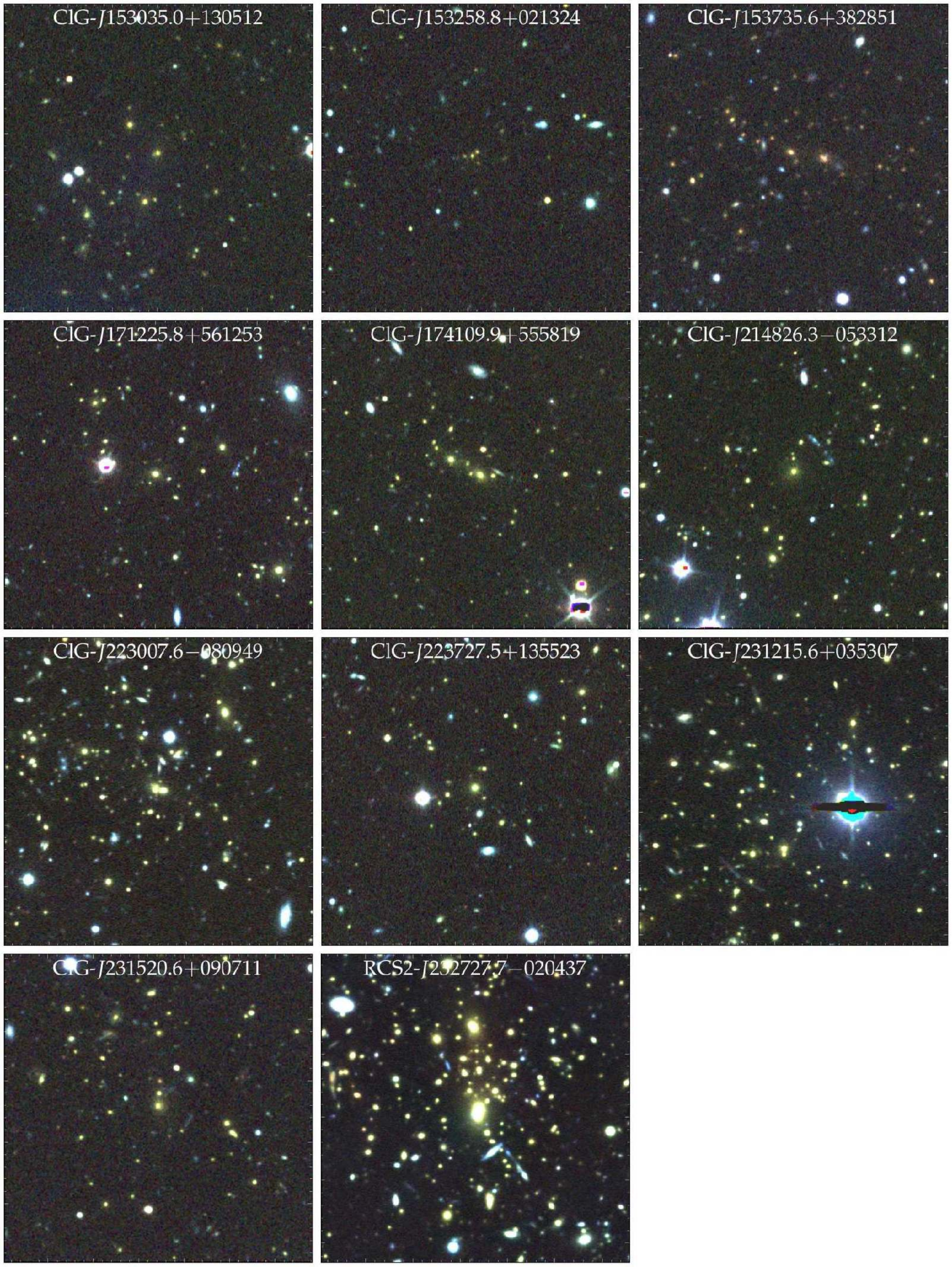}
 \contcaption{}
 \label{fig:opt_all4}
\end{figure*}

\section{Maps from the CARMA Data}

\begin{figure*}
 \centering
 \includegraphics[width=18cm,height=30cm,keepaspectratio=true]{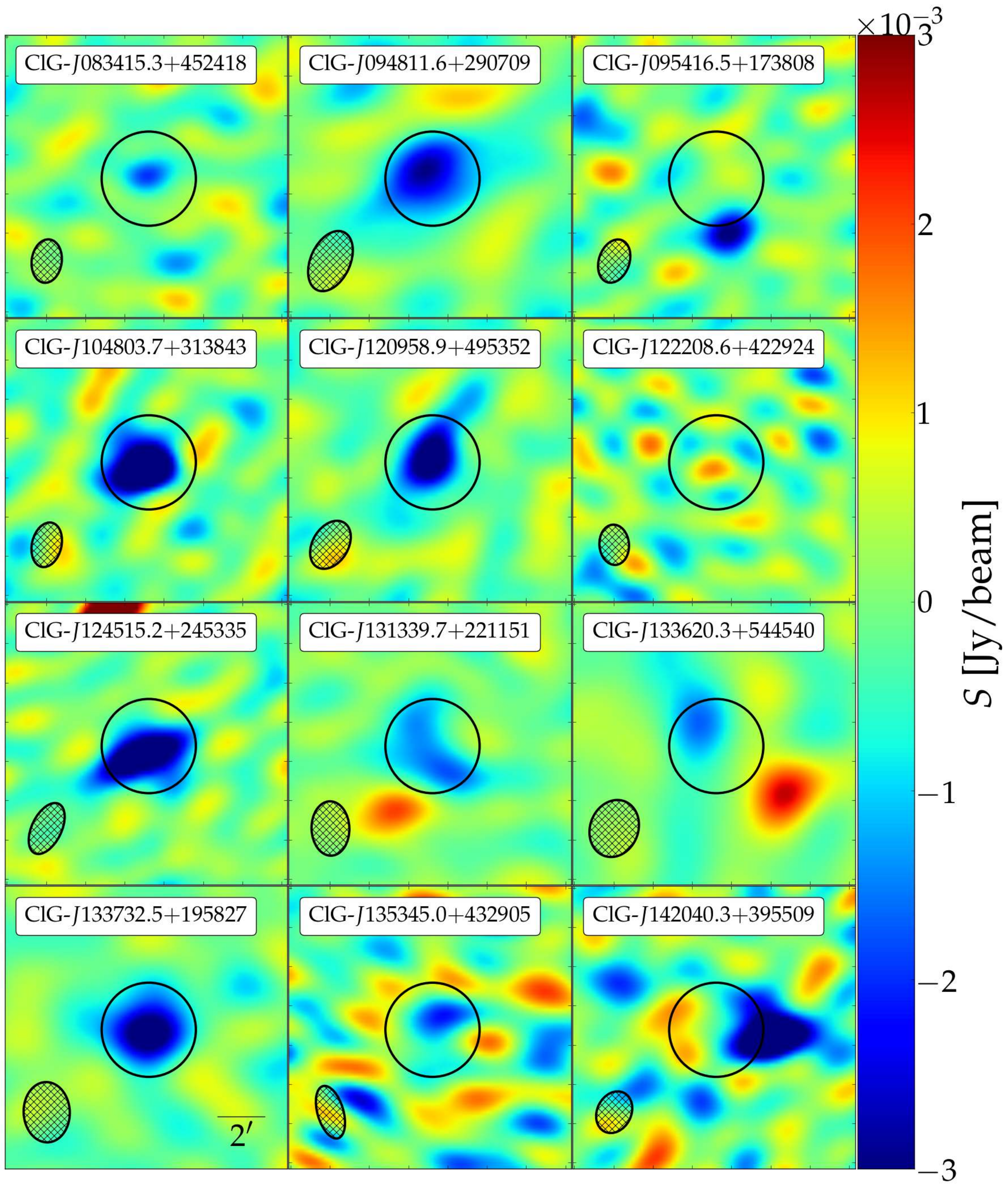}
 \caption{We show SZ-maps for all clusters observed with CARMA. The images show $12\times 12\arcmin$. 
          The ellipses in the bottom left are the beams, the circle in the centre has a $2\arcmin$ 
          radius and indicates the BCG position. }
 \label{fig:carma1}
\end{figure*}

\begin{figure*}
 \centering
 \includegraphics[width=18cm,height=30cm,keepaspectratio=true]{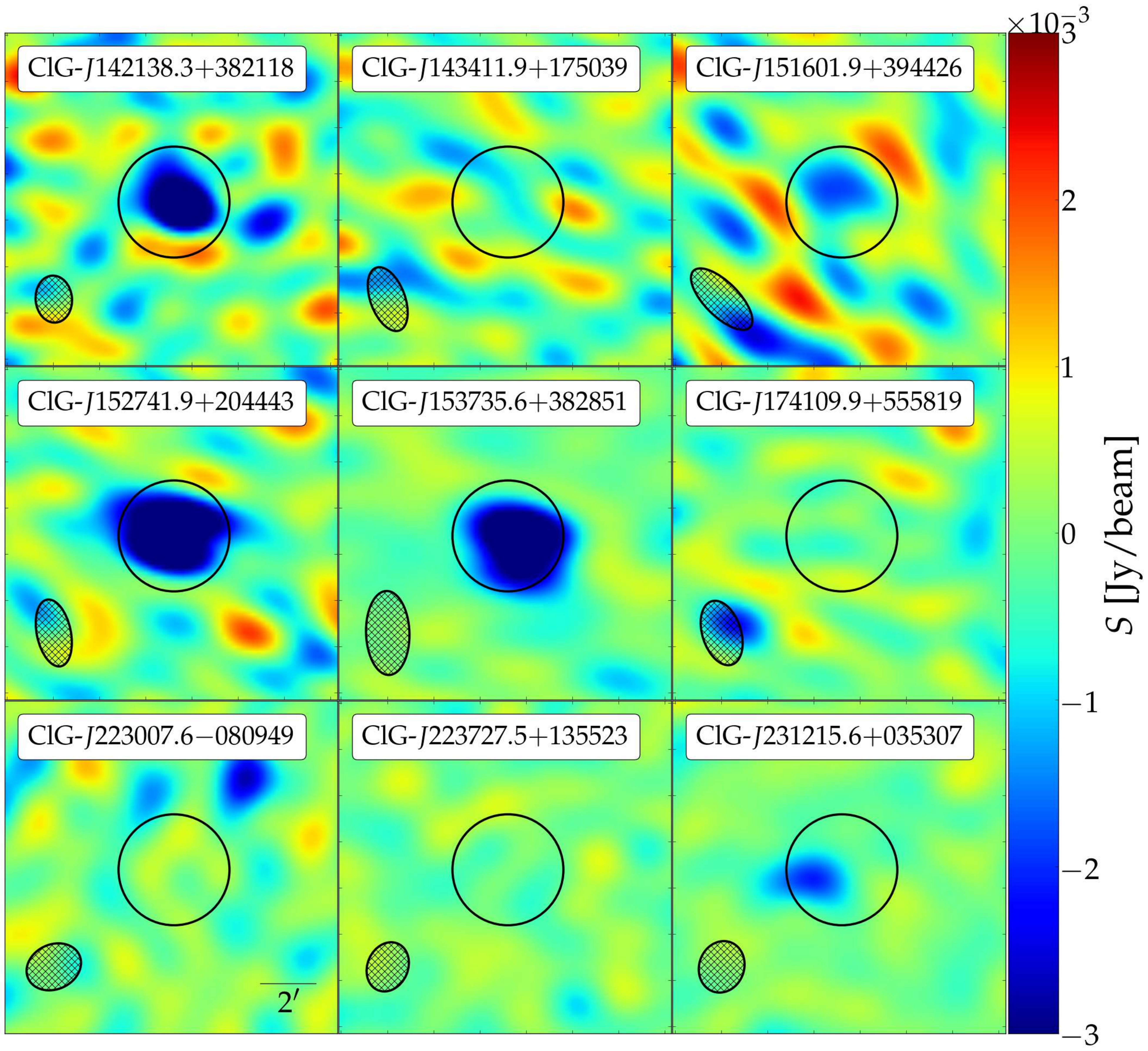}
 \contcaption{}
 \label{fig:carma1}
\end{figure*}

\begin{figure*}
 \includegraphics[width=16cm,height=30cm,keepaspectratio=true]{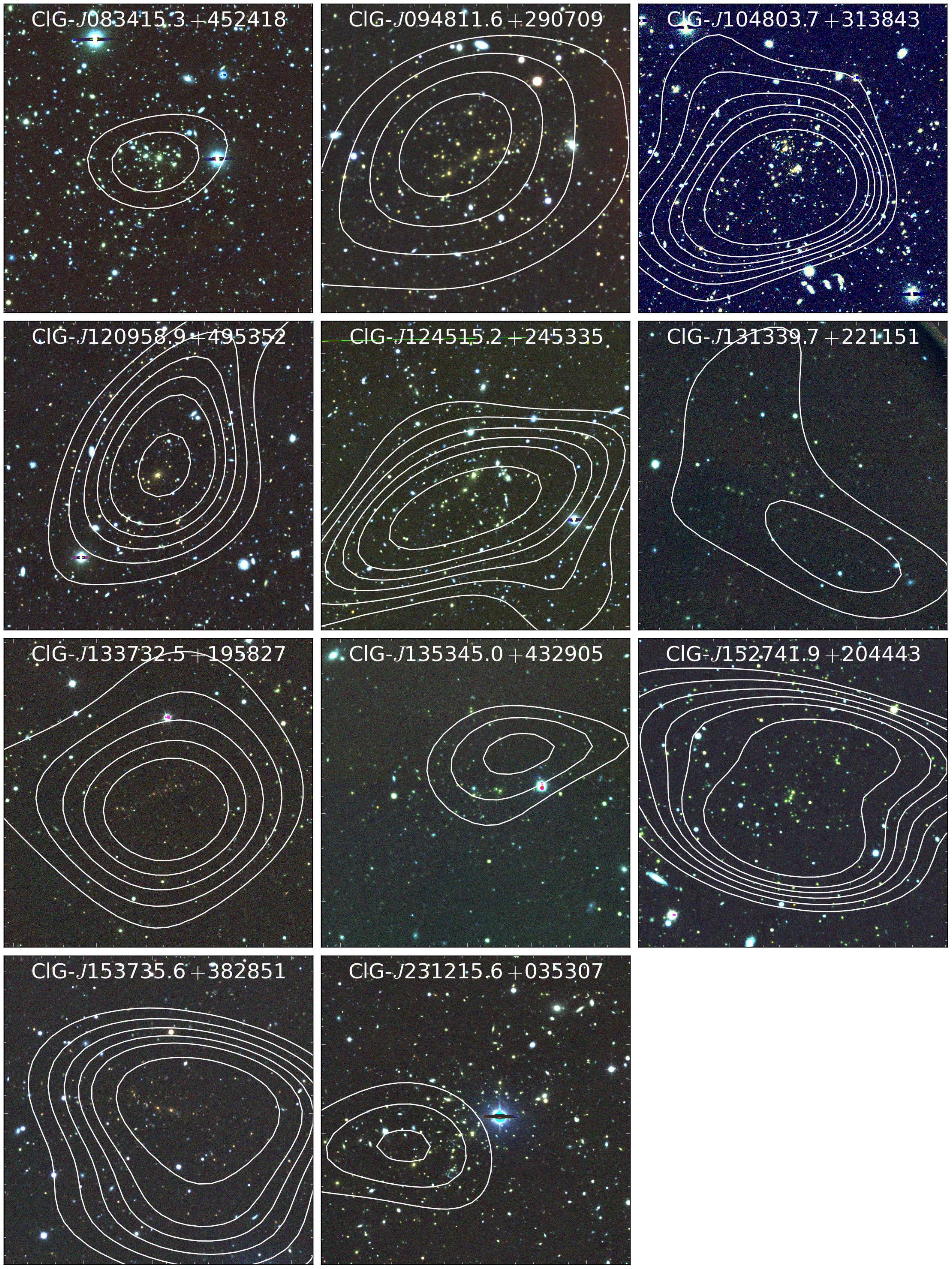}
 \caption{We show optical three-colour images and the corresponding SZ-overlay for all clusters that 
          have been detected at more than $3\sigma$ with CARMA. 
          The images show $4\farcm2\times 4\farcm2$ around the BCG. The contour levels are 
          $-4.0,-3.0,-2.5,-2.0,-1.5$ and $-1.0\times10^{-3}\rm{Jy/beam}$.}
 \label{fig:overlay}
\end{figure*}

\section{Results from \textit{Planck} Data}
\label{sec:planck}

We constructed $10^{\circ} \times 10^{\circ}$ $y$-maps of all 44 galaxy clusters from public \textit{Planck} 
data by forming a linear combination of maps (the ILC method; \citealt{sz_maps2}; \citealt{sz_maps}),
using all six frequency bands of the \textit{Planck} High Frequency Instrument (HFI), 
\changestwo{taken from the recent 2015 data release \changesthree{\citep{planck_data}}. }

A Gaussian filter was applied to smooth all maps to a common resolution of $10\arcmin$, corresponding 
to the \textit{Planck} beam at $100 \, {\rm GHz}$. The final Compton-$y$ maps are the weighted sum of all six maps: 
$y = \sum_i \omega_i T_i / T_{\mathrm{CMB}}$. Here $T_i$ are the individual channel maps, each weighted with an 
ILC-coefficient $\omega_i$.
The coefficients are chosen to minimize the variance of the reconstructed Compton-$y$ map while fulfilling two 
constrains: (1) eliminate the primary CMB Temperature anisotropies and (2) preserve the temperature fluctuations 
introduced by the SZE. The produced map may contain an offset, since the variance of the map stays unaffected 
while adding a constant. The map offset was determined by fitting the histogram of pixel values with a Gaussian, 
which provides a very good model for the map noise. The offset is then corrected for by subtracting the mean of 
the Gaussian.
In Fig. \ref{fig:sz_planck} we show \textit{Planck} $y$-maps for all clusters, which have been observed with CARMA.

\begin{figure*}
 \centering
 \includegraphics[width=18cm,height=30cm,keepaspectratio=true]{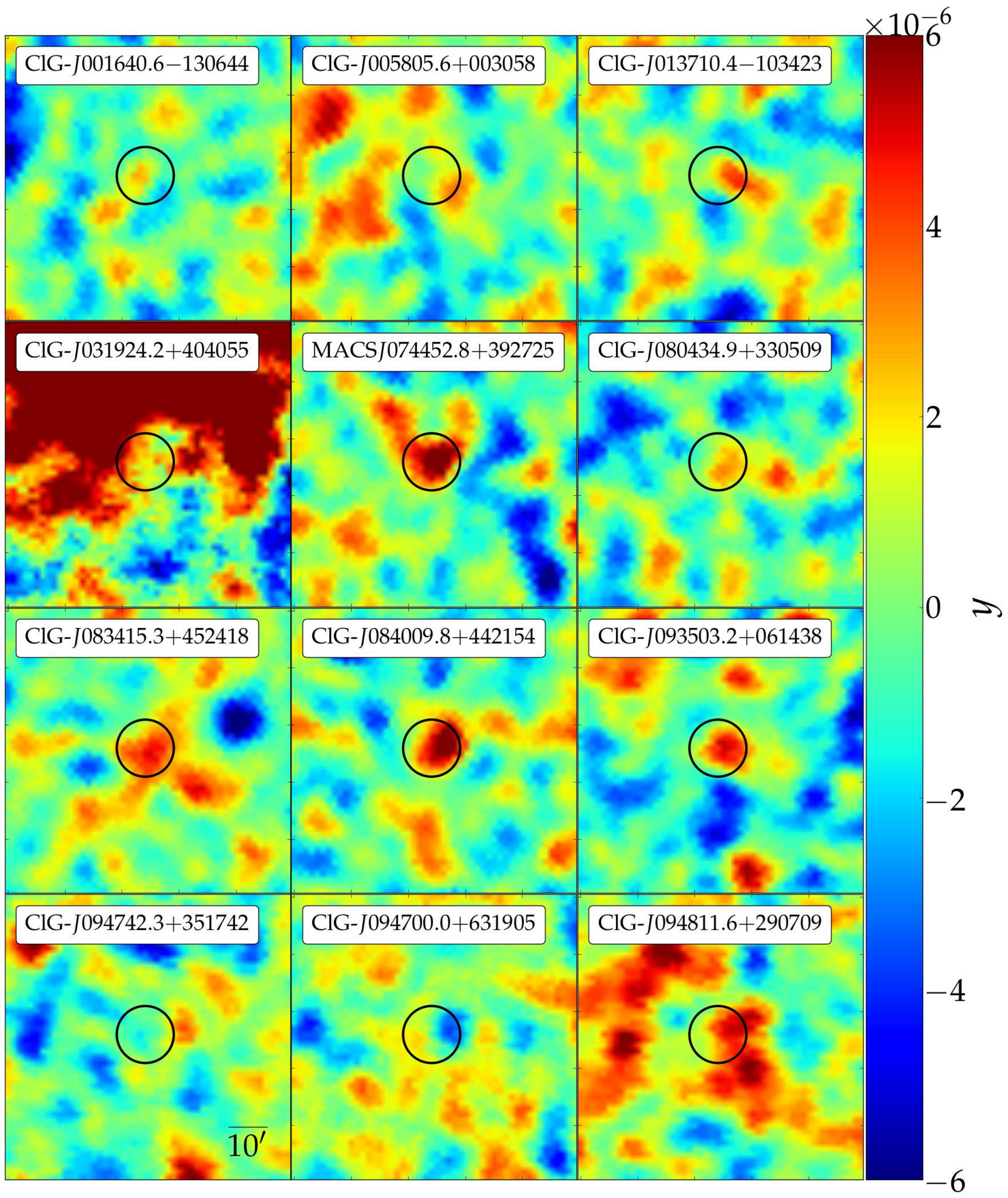}
 \caption{\changes{This figure shows the $y$-maps of all the clusters in our sample in \textit{Planck}}
          The images show a $1.25\times1.25\,\rm{deg^{2}}$ field
          around the cluster. The black circle has a $7\farcm5$ radius and is centred at the BCG. }
 \label{fig:sz_planck}
\end{figure*}

\begin{figure*}
 \centering
 \includegraphics[width=18cm,height=30cm,keepaspectratio=true]{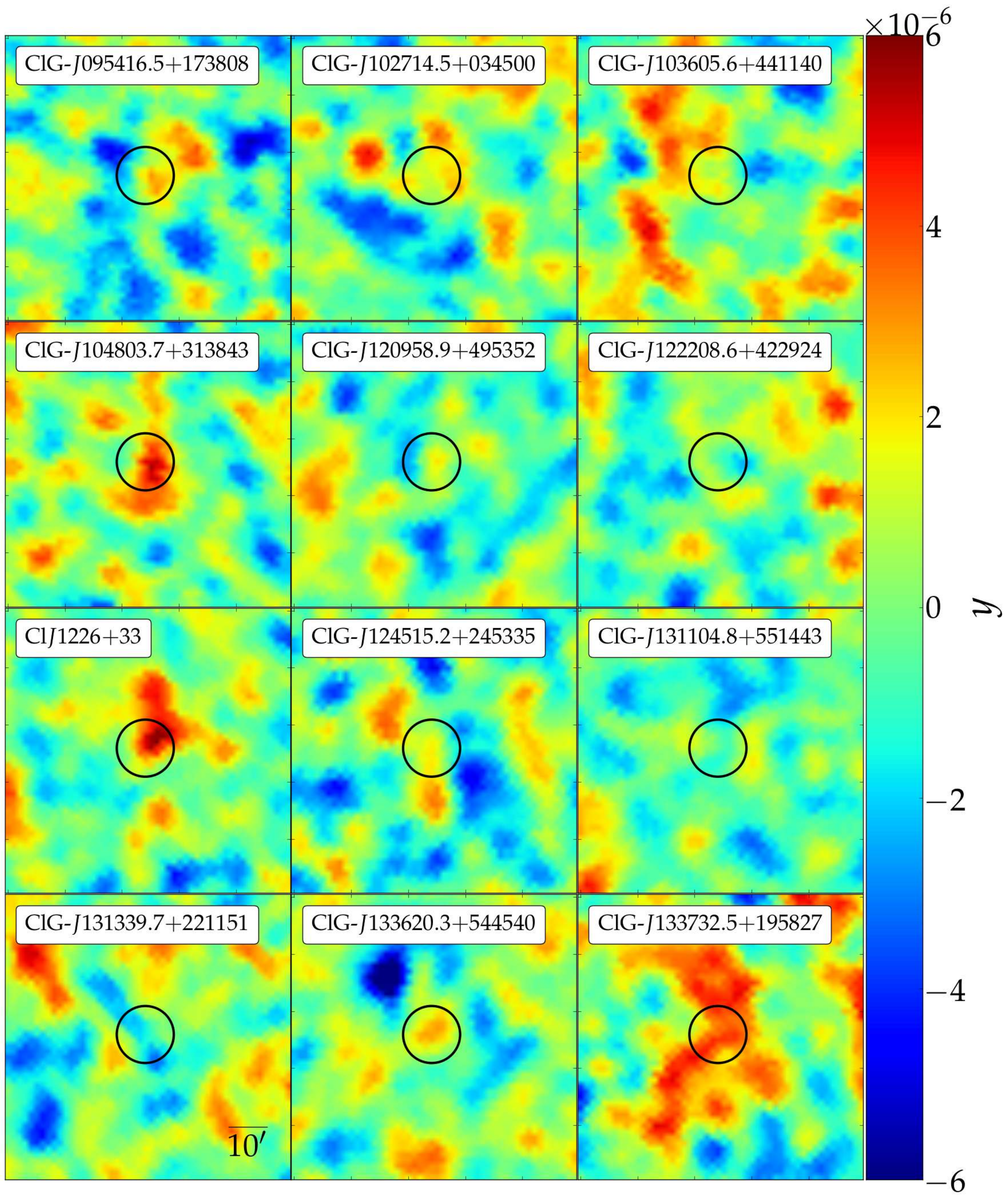}
 \contcaption{}
 \label{fig:sz_planck}
\end{figure*}

\begin{figure*}
 \centering
 \includegraphics[width=18cm,height=30cm,keepaspectratio=true]{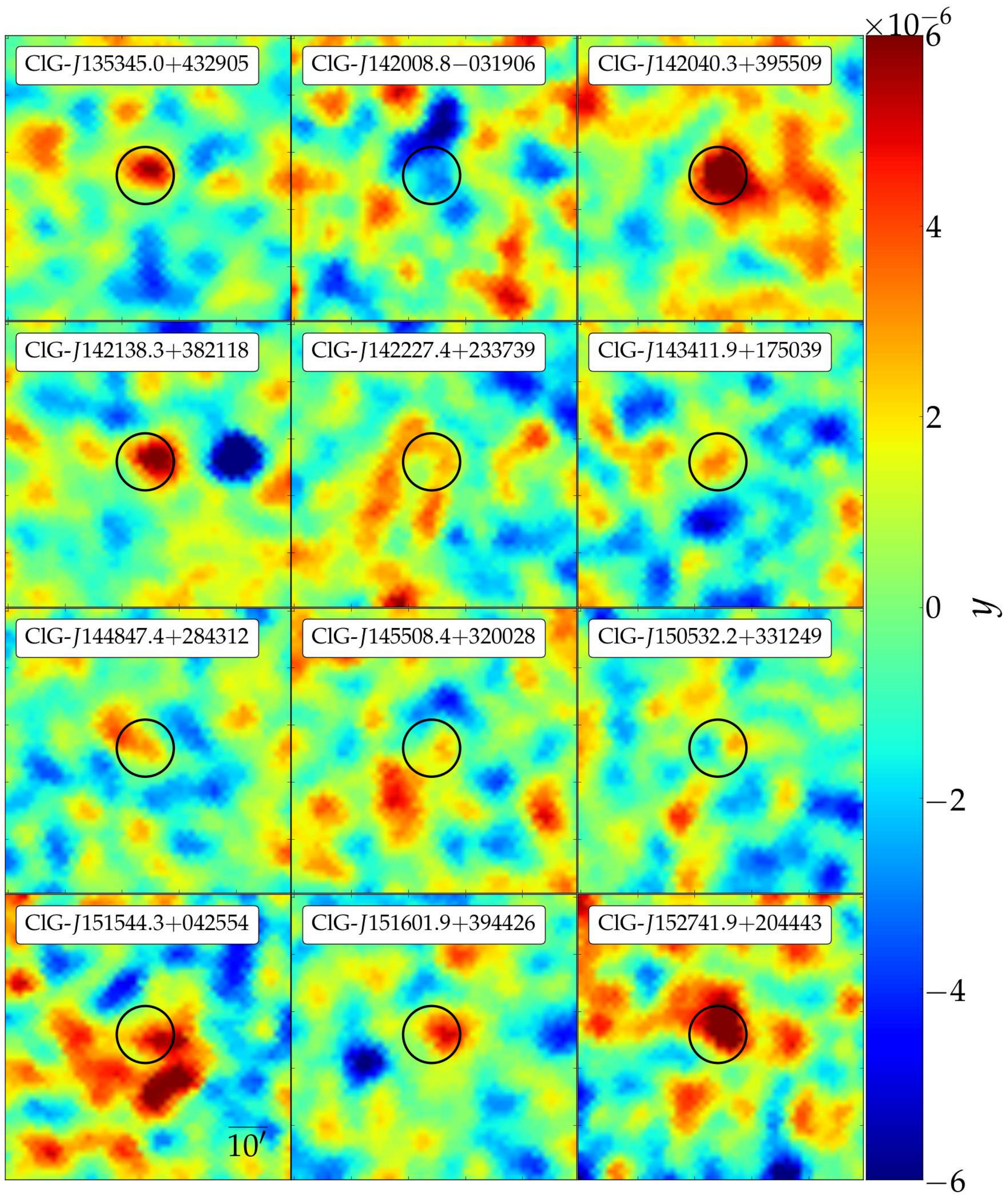}
 \contcaption{}
 \label{fig:sz_planck}
\end{figure*}

\begin{figure*}
 \centering
 \includegraphics[width=18cm,height=30cm,keepaspectratio=true]{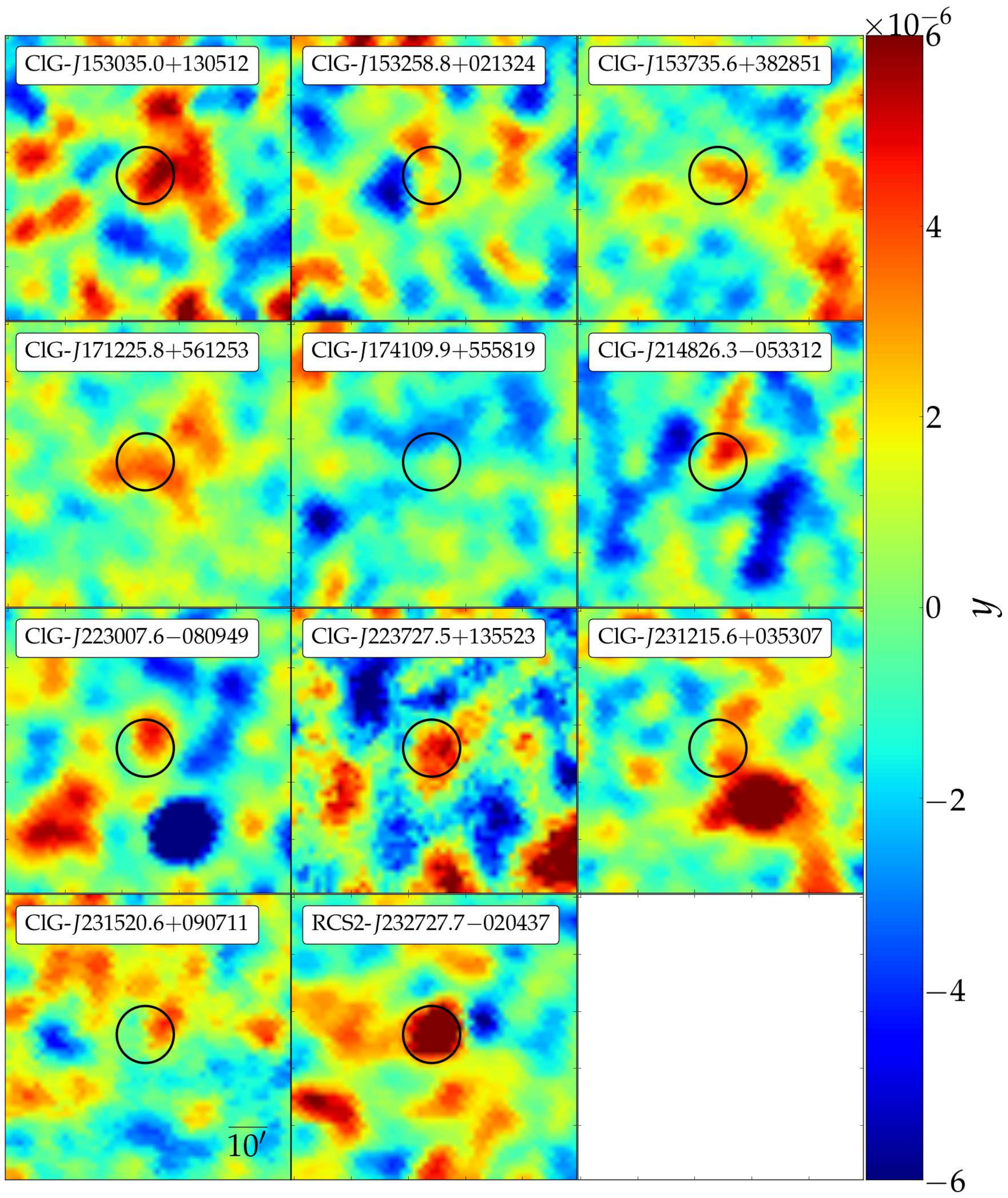}
 \contcaption{}
 \label{fig:sz_planck}
\end{figure*}

\end{document}